\documentclass[aps,prc,twocolumn,superscriptaddress,showpacs,showkeys,amsmath,amssymb,floatfix]{revtex4}
\usepackage{booktabs,graphicx,mathrsfs,verbatim}
\usepackage[usenames,dvipsnames]{color}
\usepackage{multirow,bigdelim}
\usepackage{ulem}
\usepackage{color}
\usepackage{textcomp} 
\usepackage[sort&compress]{natbib} 
\newcommand{\valencia}{\affiliation{Instituto de F{\'i}sica Corpuscular, CSIC-Universidad de Valencia, E-46071 Valencia, Spain}}
\newcommand{\chile}{\affiliation{Comisi{\'o}n Chilena de Energ{\'i}a Nuclear, Casilla 188-D, Santiago, Chile}}
\newcommand{\debrecen}{\affiliation{Inst. of Nuclear Research of the Hung. Acad. of Sciences, Debrecen, H-4026, Hungary}}
\newcommand{\surrey}{\affiliation{Department of Physics, University of Surrey, Guildford GU2 7XH, Surrey, UK}}
\newcommand{\bordeaux}{\affiliation{Centre d'Etudes Nucl{\'e}aires de Bordeaux Gradignan, CNRS/IN2P3 - Universit{\'e} de Bordeaux, 33175 Gradignan Cedex, France}}
\newcommand{\tokyo}{\affiliation{Department of Natural Sciences, Tokyo City University, 1-28-1 Tamazutsumi, Setagaya-ku, Tokyo 158-8557, Japan}}
\newcommand{\riken}{\affiliation{RIKEN Nishina Center, 2-1 Hirosawa, Wako, Saitama 351-0198, Japan}}
\newcommand{\osakadep}{\affiliation{Department of Physics, Osaka University, Toyonaka, Osaka 560-0043, Japan}}
\newcommand{\koeln}{\affiliation{Institute of Nuclear Physics, University of Cologne, D-50937 Cologne, Germany}}
\newcommand{\munich}{\affiliation{Physik Department E12, Technische Universit{\"a}t M{\"u}nchen, D-85748 Garching, Germany}}
\newcommand{\lnl}{\affiliation{Laboratori Nazionali di Legnaro INFN, I-35020 Legnaro, Padova, Italy}}
\newcommand{\bucharest}{\affiliation{National Institute for Physics and Nuclear Engineering IFIN-HH, P.O. Box MG-6, Bucharest-Magurele, Romania}}
\newcommand{\padova}{\affiliation{INFN Sezione di Padova and Dipartimento di Fisica, Universit{\'a} di Padova, I-35131 Padova, Italy}}
\newcommand{\instanbul}{\affiliation{Department of Physics, Istanbul University, Istanbul, 34134, Turkey}}
\newcommand{\caen}{\affiliation{Grand Acc{\'e}l{\'e}rateur National d'Ions Lourds (GANIL), CEA/DRF-CNRS/IN2P3, Bvd Henri Becquerel, 14076 Caen, France}}
\newcommand{\tennessee}{\affiliation{Department of Physics and Astronomy, University of Tennessee, Knoxville, Tennessee 37996-1200, USA}}
\newcommand{\tokyoII}{\affiliation{Department of Physics, Tokyo University of Science, Noda, Chiba 278-8510, Japan}}

\begin{document} 

\title{Beta decay of the very neutron-deficient $^{60}$Ge and $^{62}$Ge nuclei}

\author{S.~E.~A.~Orrigo}\email{sonja.orrigo@ific.uv.es}\valencia
\author{B.~Rubio}\valencia
\author{W.~Gelletly}\valencia\surrey
\author{P.~Aguilera}\valencia\chile
\author{A.~Algora}\valencia\debrecen
\author{A.~I.~Morales}\valencia
\author{J.~Agramunt}\valencia
\author{D.~S.~Ahn}\riken
\author{P.~Ascher}\bordeaux
\author{B.~Blank}\bordeaux
\author{C.~Borcea}\bucharest
\author{A.~Boso}\padova
\author{R.~B.~Cakirli}\instanbul
\author{J.~Chiba}\tokyoII
\author{G.~de~Angelis}\lnl
\author{G.~de~France}\caen
\author{F.~Diel}\koeln
\author{P.~Doornenbal}\riken
\author{Y.~Fujita}\osakadep
\author{N.~Fukuda}\riken
\author{E.~Ganio{\u{g}}lu}\instanbul
\author{M.~Gerbaux}\bordeaux
\author{J.~Giovinazzo}\bordeaux
\author{S.~Go}\tennessee
\author{T.~Goigoux}\bordeaux
\author{S.~Gr{\'e}vy}\bordeaux
\author{V.~Guadilla}\valencia
\author{N.~Inabe}\riken
\author{G.~Kiss}\riken
\author{T.~Kubo}\riken
\author{S.~Kubono}\riken
\author{T.~Kurtukian-Nieto}\bordeaux
\author{D.~Lubos}\munich
\author{C.~Magron}\bordeaux
\author{F.~Molina}\chile
\author{A.~Montaner-Piz{\'a}}\valencia
\author{D.~Napoli}\lnl
\author{D.~Nishimura}\tokyo
\author{S.~Nishimura}\riken
\author{H.~Oikawa}\tokyoII
\author{Y.~Shimizu}\riken
\author{C.~Sidong}\riken
\author{P.-A.~S{\"o}derstr{\"o}m}\riken
\author{T.~Sumikama}\riken
\author{H.~Suzuki}\riken
\author{H.~Takeda}\riken
\author{Y.~Takei}\tokyoII
\author{M.~Tanaka}\osakadep
\author{P.~Vi}\riken
\author{J.~Wu}\riken
\author{S.~Yagi}\tokyoII
%

\begin{abstract}
We report here the results of a study of the $\beta$ decay of the proton-rich Ge isotopes, $^{60}$Ge and $^{62}$Ge, produced in an experiment at the RIKEN Nishina Center. We have improved our knowledge of the half-lives of $^{62}$Ge (73.5(1) ms), $^{60}$Ge (25.0(3) ms) and its daughter nucleus, $^{60}$Ga (69.4(2) ms). We measured individual $\beta$-delayed proton and $\gamma$ emissions and their related branching ratios. Decay schemes and absolute Fermi and Gamow-Teller transition strengths have been determined. The mass excesses of the nuclei under study have been deduced. A total $\beta$-delayed proton-emission branching ratio of 67(3)\% has been obtained for $^{60}$Ge. New information has been obtained on the energy levels populated in $^{60}$Ga and on the 1/2$^-$ excited state in the $\beta p$ daughter $^{59}$Zn. We extracted a ground state to ground state feeding of 85.3(3)\% for the decay of $^{62}$Ge. Eight new $\gamma$ lines have been added to the de-excitation of levels populated in the $^{62}$Ga daughter.
\end{abstract}



\maketitle

\section{\label{intro}Introduction}

The investigation of nuclear structure close to the limits of nuclear existence is one of the frontiers of modern Nuclear Physics. The study of the properties of exotic nuclei is crucial to provide tests of the predictions of nuclear models at extreme values of isospin. The experimental challenge involved with the production of such unstable nuclei has resulted in a worldwide effort to build next-generation facilities producing and accelerating radioactive ion beams (RIBs) of very high intensity. Heavy neutron-deficient nuclei can now be produced up to the proton drip-line, enabling the realization of detailed decay studies and the observation of new exotic decay modes \cite{Blank2008,Pfutzner2012,PhysRevLett.112.222501,PhysRevC.93.044336}.

The structural properties of exotic proton-rich nuclei are also important for Nuclear Astrophysics, because many of them lie on the $rp$-process (rapid proton-capture) reaction pathway leading to the production of the heavy elements in the Universe in explosive stellar environments \cite{Schatz1998167,Fisker2008,Parikh2013225,Galloway2004}. 

Decay spectroscopy experiments with implanted RIBs are a powerful tool to explore the properties of exotic nuclei and provide rich spectroscopic information: $\beta$-decay half-lives, delayed emission of $\gamma$ rays, particle-decay branching ratios (commonly protons are emitted in the case of proton-rich nuclei), excited states populated in the daughter nucleus, and so on. Furthermore $\beta$ decay provides direct access to the absolute values of the Fermi $B$(F) and Gamow-Teller $B$(GT) transition strengths. 

In this paper we present new results on the $\beta$ decay of the neutron-deficient Ge isotopes, $^{62}$Ge and $^{60}$Ge. They were produced and implanted in unprecedented numbers at the Radioactive Ion Beam Factory (RIBF) of the RIKEN Nishina Center (Japan) thanks to the availability of a high-intensity beam of $^{78}$Kr. $^{62}$Ge is a $T_z=\text{-}1$ nucleus about which little was known at the time of our experiment. $^{60}$Ge is a $T_z=\text{-}2$ nucleus whose decay is almost unknown, apart from a first measurement of its half-life with 28 events \cite{ciemny2016}. $^{60}$Ge is a semi-magic \mbox{$N$ = 28} isotone and the heaviest one for which the mirror $^{60}$Ni nucleus is stable, allowing one to explore mirror symmetry. The $^{60}$Ga daughter lies right at the proton drip-line and nothing is known of its level scheme. Hence our study provides brand new information.

In a more general context, this work is part of a systematic study of proton-rich nuclei which we have carried out at different RIB facilities \cite{PhysRevLett.112.222501,PhysRevC.93.044336,PhysRevC.94.044315,Kucuk2017,PhysRevC.91.014301}, also in comparison with mirror charge-exchange experiments done on the mirror stable target \cite{Fujita2011549}. One focus of interest for the $T_z=\text{-}2$ nuclei such as $^{60}$Ge is to explore the competition between the $\gamma$ de-excitation and (isospin-forbidden) proton emission from the $T$ = 2 Isobaric Analogue State (IAS), that is populated in the daughter nucleus by the $\beta$ decay. This feature has been observed in all the lighter $T_z=\text{-}2$ systems already studied \cite{PhysRevLett.112.222501,PhysRevC.93.044336}. The $T_z=\text{-}1$ nuclei such as $^{62}$Ge are also of interest to check whether the suppression of the isoscalar $\gamma$ transitions between \mbox{$J^{\pi}$ = 1$^+$}, $T$ = 0 states (Warburton and Weneser $quasi$-$rule$ \cite{Mompurgo58,Wilkinson69}) observed in previous $T_z=\text{-}1$ cases \cite{PhysRevC.91.014301} persists.

The paper is organized as follows. Section \ref{exp} describes the experiment and the setup. The details of the data analysis are described in Section \ref{an}. The results obtained on the $\beta^+$ decay of $^{60}$Ge are given in Section \ref{ge60}, together with new results on the $\beta^+$ decay of the $^{60}$Ga daughter. Section \ref{ge62} shows the results on the $\beta^+$ decay of $^{62}$Ge. Finally, Section \ref{concl} summarises the conclusions.

\section{\label{exp}The experiment}

\begin{figure}[!t]
  \centering
	\includegraphics[width=1\columnwidth]{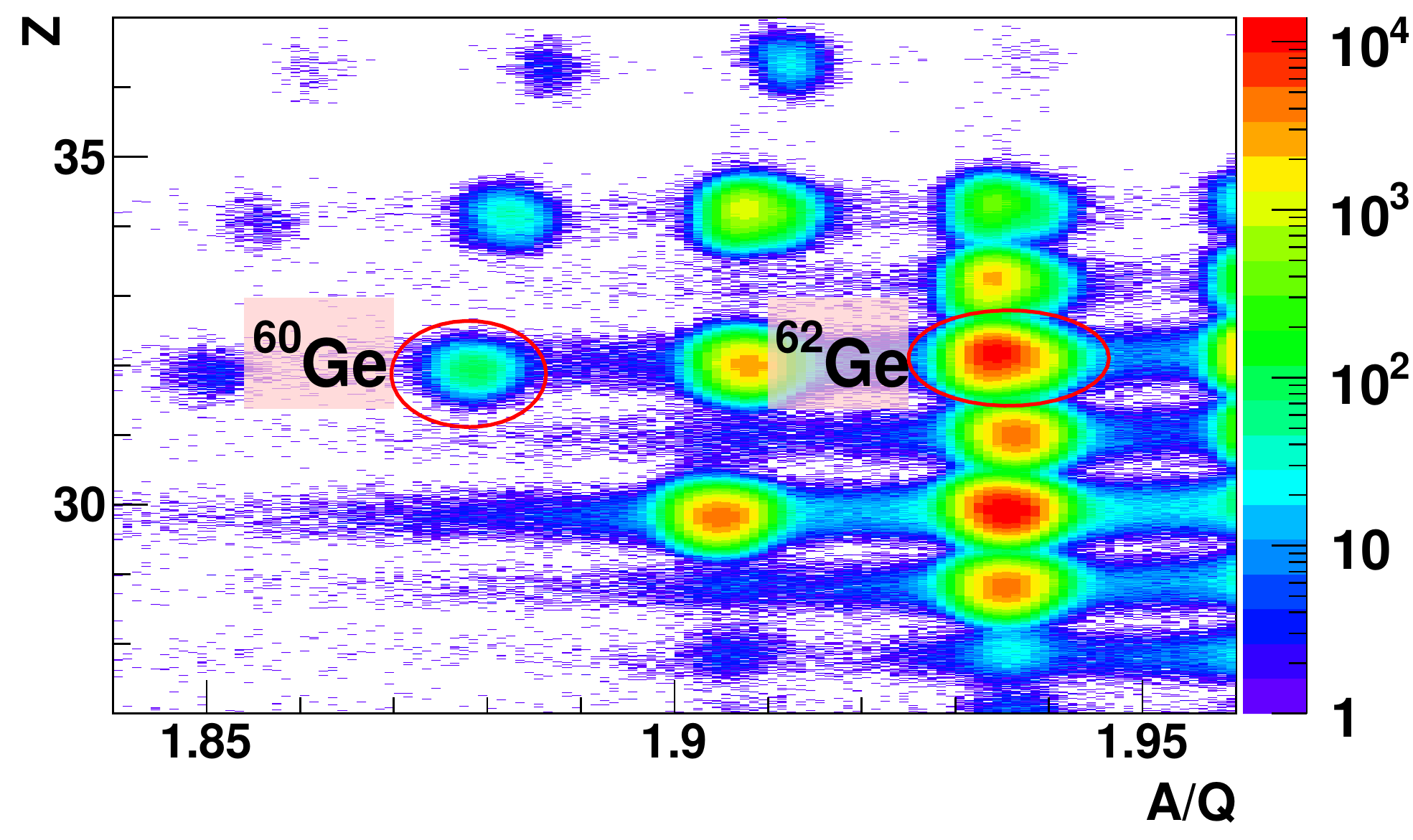}
 	\caption{$Z$ versus $A/Q$ identification plot for the BigRIPS setting optimized for $^{65}$Br, in which both $^{60}$Ge and $^{62}$Ge are implanted. The positions of these nuclei are shown.}
  \label{IDplot}
\end{figure}

The $^{60}$Ge and $^{62}$Ge proton-rich nuclei were produced with unprecedented intensity in an experiment performed at RIBF, RIKEN Nishina Center (Japan). A high-intensity (up to 250 pnA) $^{78}$Kr primary beam was accelerated to 345 MeV/nucleon and fragmented on a $^9$Be target with thickness of 5 mm. The exotic fragments produced in this way were separated, selected and identified in the BigRIPS separator \cite{Kubo2003,Kubo2012} by means of the $B\rho - \Delta E - ToF$ method \cite{Fukuda2013}. A series of parallel-plate avalanche counters (PPAC), multisampling ionization chambers (MUSIC) and plastic scintillators constitute the detection setup of BigRIPS and are employed to measure the position of the transmitted ions at different focal planes (which is related to the magnetic rigidity $B\rho$), their energy loss $\Delta E$ and time of flight $ToF$, respectively. Standard particle-identification procedures, together with additional off-line cuts on the BigRIPS variables to remove background events thoroughly, allow one to identify the ions by their atomic number $Z$ and mass-to-charge ratio $A/Q$ with high resolution \cite{Fukuda2013,Blank2016}. An example of the two-dimensional identification matrix obtained for the BigRIPS setting optimized for $^{65}$Br is shown in Fig. \ref{IDplot}, where the positions of $^{60}$Ge and $^{62}$Ge are indicated. It should be noted that the odd $^{59}$Ge and $^{61}$Ge isotopes were also produced and their decay was studied in Refs. \cite{Blank2016,Goigoux2016}.
  
The selected and identified fragments were then transmitted to the exit of the ZeroDegree spectrometer (ZDS) \cite{Kubo2012} where a setup for $\beta$-decay spectroscopy was installed. This setup consisted of the Wide-range Active Silicon Strip Stopper Array for Beta and ion detection (WAS3ABi) \cite{Nishimura2012} and the EUroball-RIKEN Cluster Array (EURICA) \cite{Soderstrom2013}. A common timestamp with a frequency of 10$^8$ Hz was used to correlate the data acquisition systems of BigRIPS, WAS3ABi and EURICA.

The WAS3ABi array was used to detect both the implanted heavy-ions and subsequent charged-particle ($\beta$ particles and protons) decays, measuring their energy, position and time. WAS3ABi consisted of three 1-mm-thick Double-Sided Silicon-Strip Detectors (DSSSD), each one having a 6$\times$4 cm$^2$ area. Each DSSSD was segmented into 60 vertical ($X$) and 40 horizontal ($Y$) strips with a pitch of 1 mm, defining a total of 2400 pixels. Each strip was read individually by standard analogue electronics providing energy and time signals. A plastic scintillator was placed behind WAS3ABi acting as a veto, i.e., tagging the fragments that were not stopped inside WAS3ABi and also those undergoing secondary reactions in WAS3ABi.

In order to optimize the resolution for the $\beta$ particles, the gain was set to achieve a full energy range of 4 MeV for the $X$ strips, while the energy range of the $Y$ strips was increased to 10 MeV to allow the detection of high-energy protons. Known $\beta$-delayed proton emitters ($^{57}$Zn, $^{61}$Ge and $^{65}$Se) produced during the experiment, together with conversion electrons from a $^{207}$Bi source, were used for the energy calibration of WAS3ABi \cite{Goigoux2016}.

An implantation signal in WAS3ABi is characterised by an overflow energy signal. This is because the implanted fragments reach the array with energies well above 1 GeV and, with the gain settings mentioned above, they saturate the electronics. Since it is not only the implantation strip that saturates, but also the neighbouring ones, the position ($X$,$Y$) at which the implantation took place was determined by the fastest $X$ and $Y$ strips to fire. 

WAS3ABi was surrounded by the EURICA array \cite{Soderstrom2013}, employed to detect both the prompt and $\beta$-delayed $\gamma$ rays after the implantation event and measure their energies and times. In the present experiment EURICA used a total of 84 high-purity germanium detectors, arranged in 12 clusters containing 7 Ge crystals each. The $\gamma$-ray signals were acquired up to 100 $\mu$s after WAS3ABi provided a trigger in order to allow for the detection of the decay of isomeric states. EURICA was calibrated in energy using $\gamma$ rays from $^{133}$Ba, $^{152}$Eu and $^{244}$Cm sources \cite{Soderstrom2013}. The $\gamma$ efficiency was calibrated using the $^{133}$Ba and $^{152}$Eu sources, together with known $\gamma$ rays from $^{129}$Cd and $^{132}$Sn observed in a previous experiment \cite{PhysRevC.91.054324}. The array had an absolute detection efficiency of 10\% at 662 keV \cite{Morales2017}.

\section{\label{an}The data analysis}

During the experimental campaign BigRIPS was optimised for different settings \cite{Goigoux2016, Morales2017}. In the present paper we report the analysis of data from two of them. The first was a setting optimised for $^{65}$Br, where both $^{60}$Ge and $^{62}$Ge were implanted. The second was optimised for $^{64}$Se, where only $^{62}$Ge was implanted. The WAS3ABi deadtime fraction $\tau$, determined by comparing free and accepted triggers recorded for each run, was 22(3)\% for the first setting ($^{60}$Ge) and 26(3)\% for the combination of both settings ($^{62}$Ge).

Provided that the ions were identified in BigRIPS as explained in Section \ref{exp}, an implantation event had also to simultaneously satisfying the following conditions: (a) a signal in the last fast-plastic scintillator (F11) of the BigRIPS+ZDS spectrometer setup, (b) an overflow energy signal in WAS3ABi and (c) no signal in the veto plastic behind WAS3ABi.
A $\beta$-decay event was defined as an event simultaneously satisfying the following conditions: (A) no signal in F11, (B) no overflow energy signal in the $Y$ strips of WAS3ABi and (C) an energy signal above threshold (typically 50 keV) in WAS3ABi.

Standard techniques, extensively described in Section III of Ref. \cite{PhysRevC.93.044336}, were employed to perform the data analysis, starting with the time correlations between implantation and decay events, the construction of the charged-particle decay-energy spectrum and the $\beta$-delayed $\gamma$-ray energy spectrum for decay events correlated with implantations of a given nuclear species and the determination of the Fermi and Gamow-Teller transition strengths.

In particular, in the present case, for a specified DSSSD in WAS3ABi each implantation event was correlated in time with any decay event occurring before and after it in the same pixel and in the 8 pixels surrounding it. This is because given the smaller size of the pixels, namely 1 mm$^2$, the truly-correlated decay event may also occur in one of the closest pixels. This procedure accounts for the true time correlations, but at the price of introducing many random correlations. These latter will contribute to a flat background in the correlation-time spectrum which adds to the typical decay curve described by the Bateman equations \cite{Bateman1910} and can easily be taken into account in the half-life fit \cite{PhysRevC.93.044336}. Here the time correlations were performed over a period of time of [-1 s, +1 s], where the backward-correlations part [-1 s, 0] clearly includes only random correlations and is used in the background estimation procedure. The half-life results are reported with their statistical uncertainties.

For a given nuclear species, the charged-particle decay-energy spectrum measured in WAS3ABi was obtained as the sum of the spectra from all the pixels. As in Ref. \cite{PhysRevC.93.044336} a clean decay-energy spectrum was formed by subtracting a background energy spectrum, constructed by setting a gate of [-1 s, 0] on the correlation time, from the decay-energy spectrum constructed in [0, 1 s].

The $\gamma$-ray energy spectrum correlated with implantations of a given nuclear species was obtained by summing all the Ge crystals of EURICA, producing spectra both with and without addback. It should be noted that in this paper we show those with addback. A similar background subtraction procedure was employed to obtain a clean $\gamma$-energy spectrum, using the same gates of [-1 s, 0] and [0, 1 s] on the correlation time. In addition, the $\gamma$-ray time was restricted to \mbox{[0, 800 ns]} with respect to the decay signal in WAS3ABI.

\section{\label{ge60}Beta decay of $^{60}$G\MakeLowercase{e}}

The exotic neutron-deficient $^{60}$Ge nucleus is a special system to study. It is a semi-magic \mbox{$N$ = 28} isotone and the heaviest one for which the mirror nucleus ($^{60}$Ni) is stable, which makes it easier to investigate mirror symmetry. Furthermore, the decay of $^{60}$Ge is almost unknown. Nothing is known about the level scheme of the $^{60}$Ga daughter, which is a nucleus lying right at the proton drip-line. There exists a first measurement of the half-life of $^{60}$Ge \cite{ciemny2016} with 28 implantation events (19 of them with observation of $\beta$-delayed protons). The RIKEN high-intensity $^{78}$Kr beam allowed us to achieve the unprecedented statistics of 1.5$\times$10$^4$ implants of $^{60}$Ge. Moreover, $^{60}$Ge is a $T_z=\text{-}2$ nucleus, hence its study may shed light on the possible competition between the $\beta$-delayed proton emission and $\gamma$ de-excitation from the IAS in the daughter, as already observed in the decay of lighter $T_z=\text{-}2$ systems \cite{Dossat200718,PhysRevLett.112.222501,PhysRevC.93.044336}. 

\begin{figure}[!t]
  \centering
	\includegraphics[width=1\columnwidth]{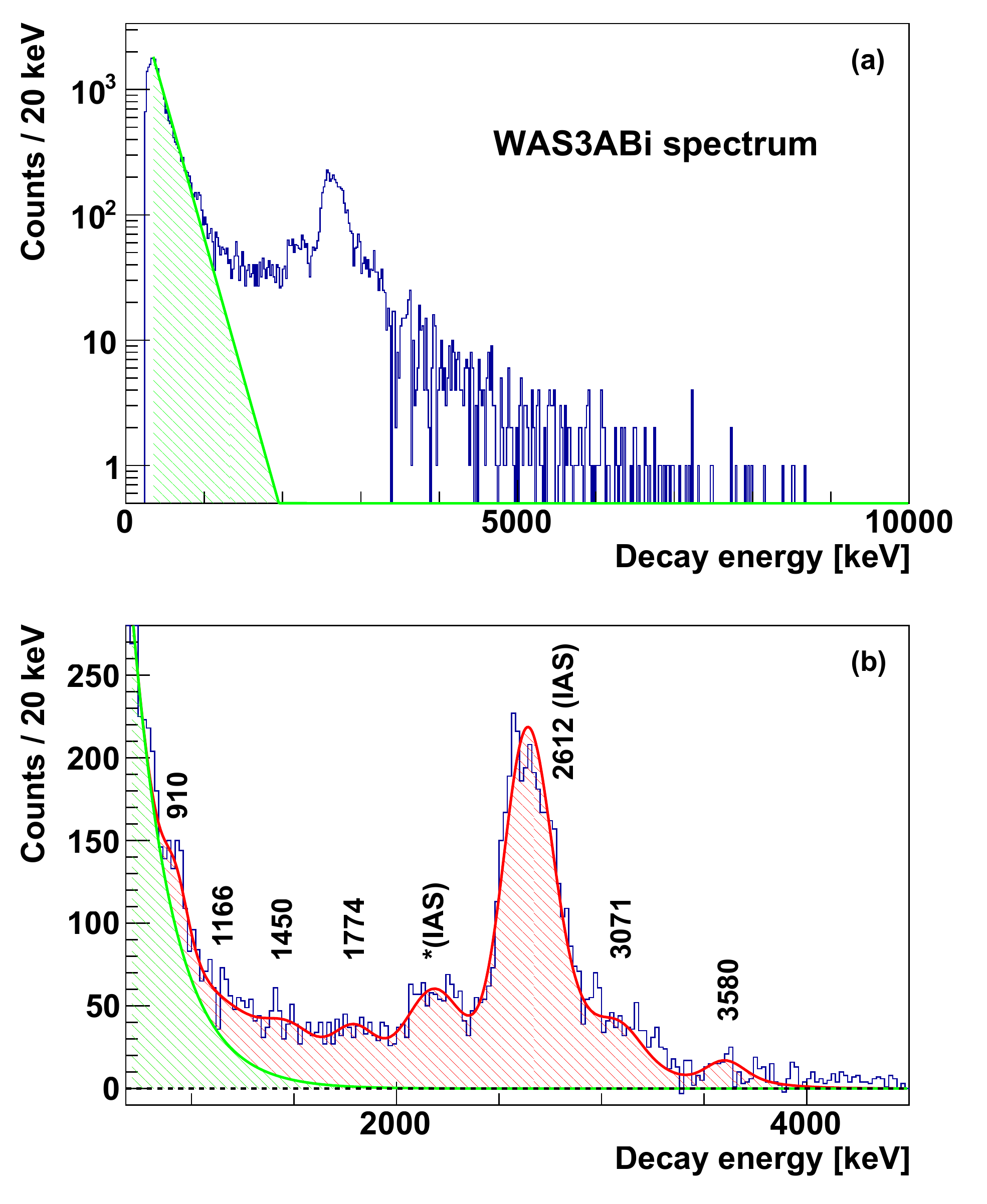}
 	\caption{(a) Charged-particle energy spectrum for decay events correlated with $^{60}$Ge implants. The peak due to the detection of $\beta$ particles is seen at low energy and fitted by an exponential function (in green). (b) Fit of the peaks related to the proton emission following the $\beta$ decay of $^{60}$Ge (in red). Peaks are labelled according to the corresponding excitation energies in the $\beta$-daughter $^{60}$Ga. The peak marked by an asterisk corresponds to the transition from the IAS to the first excited state in $^{59}$Zn (see text).}
  \label{dssd60}
\end{figure}

The charged-particle energy spectrum for decay events correlated with $^{60}$Ge implants is shown in Fig. \ref{dssd60}. The bump visible in Fig. \ref{dssd60}(a) at low energy is due to the detection of $\beta$ particles and is fitted by an exponential function (green line). Eight peaks are observed at higher energy, corresponding to $\beta$-delayed proton emission. Their fit is shown in Fig. \ref{dssd60}(b). The best candidate for the de-excitation of the IAS of $^{60}$Ge in $^{60}$Ga is the most intense peak at $E_p$ = 2522 keV which, as explained later in the text, corresponds to an excitation energy \mbox{$E_X$ = 2612} keV in $^{60}$Ga. As expected in this kind of experiment, the summing of the proton signals with those of the coincident $\beta$ particles affects the line shape of the peaks. Following our procedure \cite{PhysRevLett.112.222501,PhysRevC.93.044336}, the line shape was determined by Monte Carlo simulations and checked by fitting the well-known $^{57}$Zn peaks. The decay energies $E_p$ and their intensities $I_p$ per 100 decays are shown in Table \ref{protons}. It should be noted that $E_p$ is the total energy released in the proton decay including the energy of the recoiling nucleus.

\begin{figure}[!h]
  \centering
	\includegraphics[width=1\columnwidth]{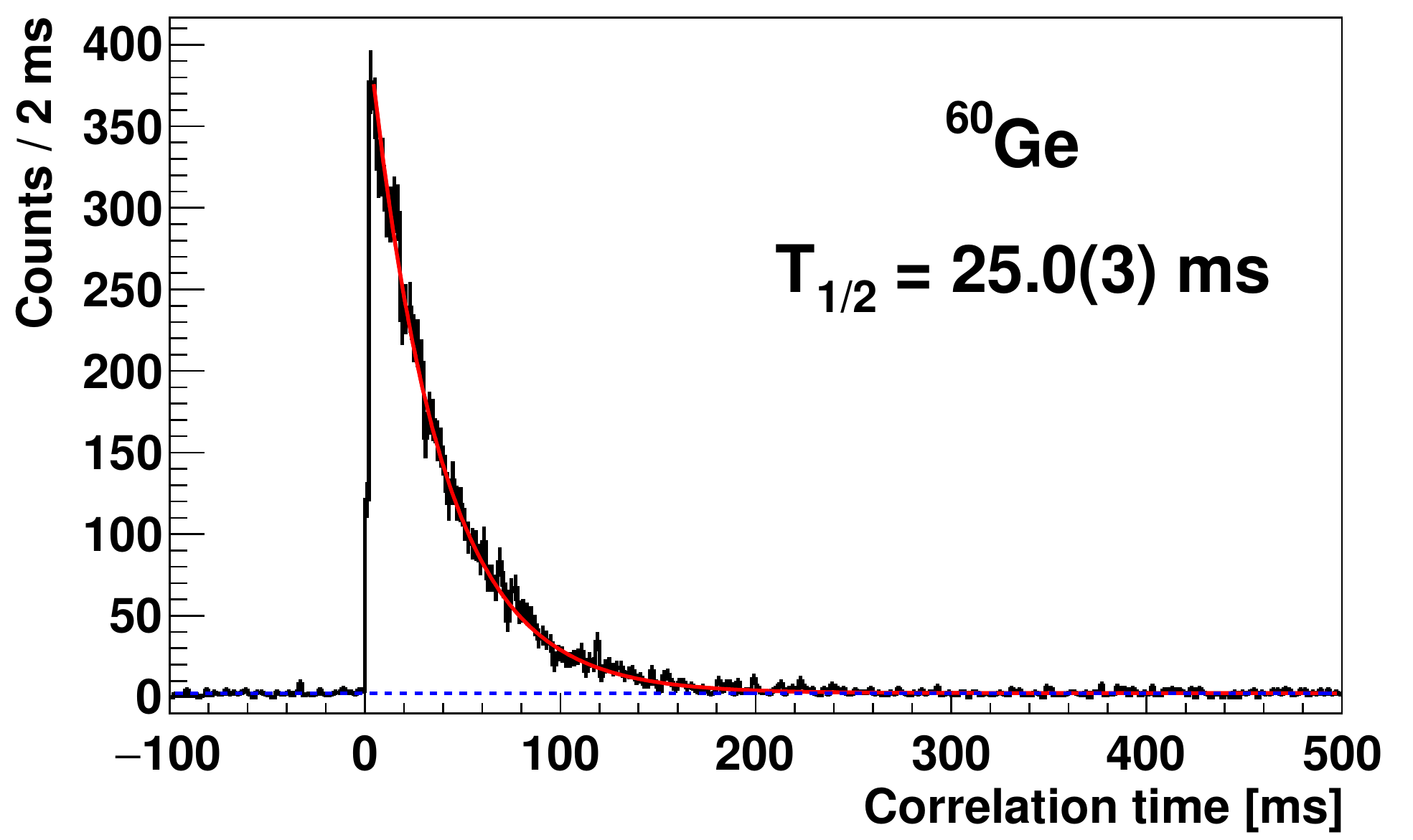}
 	\caption{Spectrum of the time correlations between each proton decay (WAS3ABi energy \mbox{$E_p>$ 1 MeV}) and all $^{60}$Ge implants. The dashed blue line is the background (fixed by fitting the backward-correlations part) and the red line is the result of the fit.}
  \label{T60}
\end{figure}

Fig. \ref{T60} shows the correlation-time spectrum obtained for $^{60}$Ge selecting only the $\beta$-delayed proton decays, by setting the condition WAS3ABi energy \mbox{$E_p>$ 1 MeV}. The half-life is determined by a least squares fit to the data including the parent activity and a linear background, fixed by a fit to the backward-correlations part, as in Ref. \cite{PhysRevC.93.044336}. A value of $T_{1/2}$ = 25.0(3) ms was obtained for $^{60}$Ge. A fit using a maximum likelihood minimization method gives a consistent result. We have improved the precision on the half-life value in comparison with the only existing measured value of 20$_{-5}^{+7}$ ms \cite{ciemny2016}.

The total proton-emission branching ratio $B_p$ is determined, as in Ref. \cite{PhysRevC.93.044336}, by comparing the total number of protons, $N_p$ (obtained from the fit shown in Fig. \ref{T60}), with the total number of implanted nuclei, $N_{imp}$, according to: 

\begin{equation}
  B_p = \frac{N_p}{N_{imp}~(1-\tau)} \,.
  \label{Eq1}
\end{equation}
where $\tau$ is the deadtime fraction (see Section \ref{an}). A value $B_p$ = 67(3)\% is obtained, where the uncertainty takes into account both the statistical and systematics ones. The latter was estimated as previously \cite{Dossat200718,PhysRevC.93.044336} by repeating the determination of $B_p$ changing the condition on the energy $E_p$ by $\pm$100 keV.

\begin{figure}[htb]
	\begin{minipage}{1.0\linewidth}
    \centering
    \includegraphics[width=1\columnwidth]{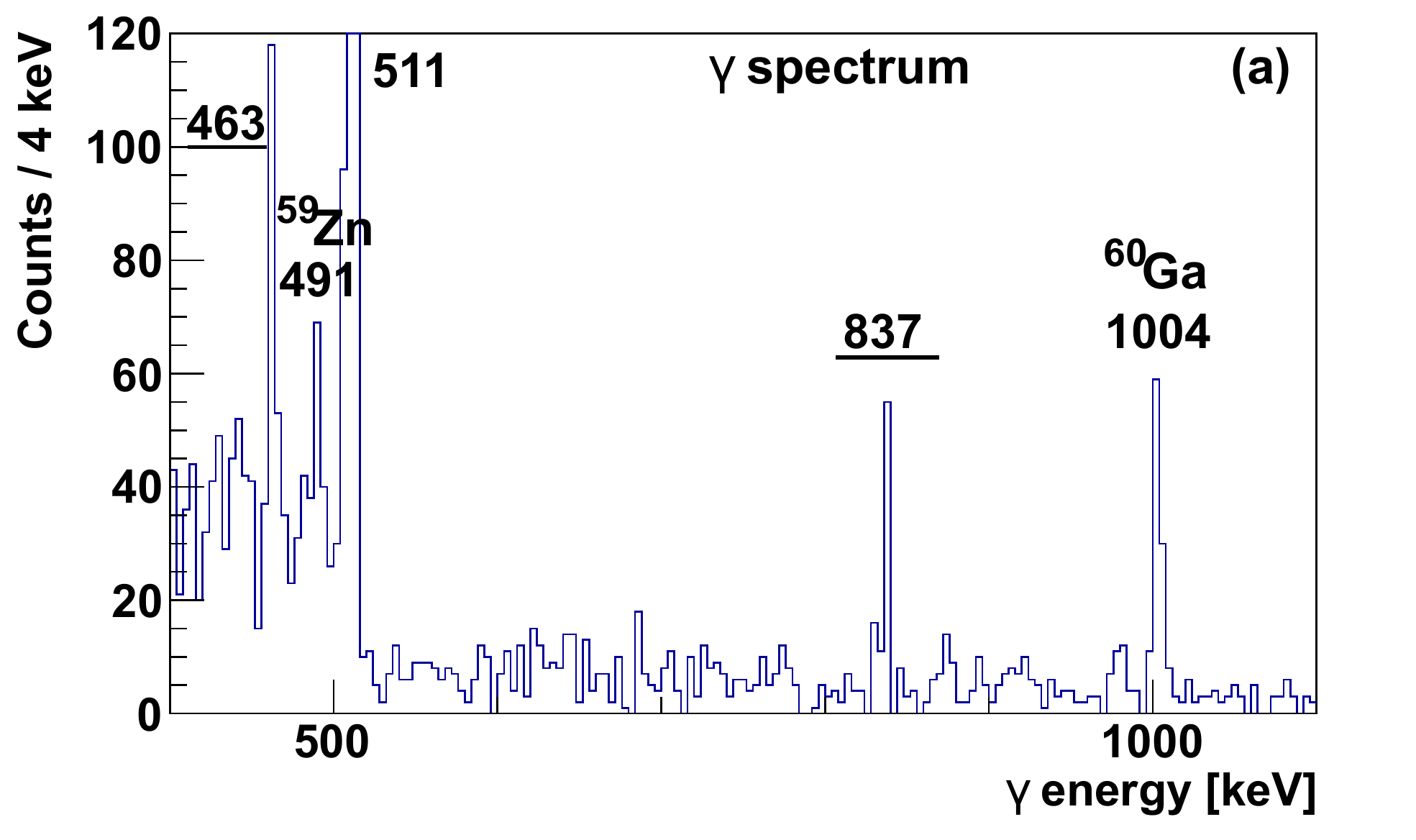}
	\end{minipage}
	\begin{minipage}{1.0\linewidth}
	  \includegraphics[width=1\columnwidth]{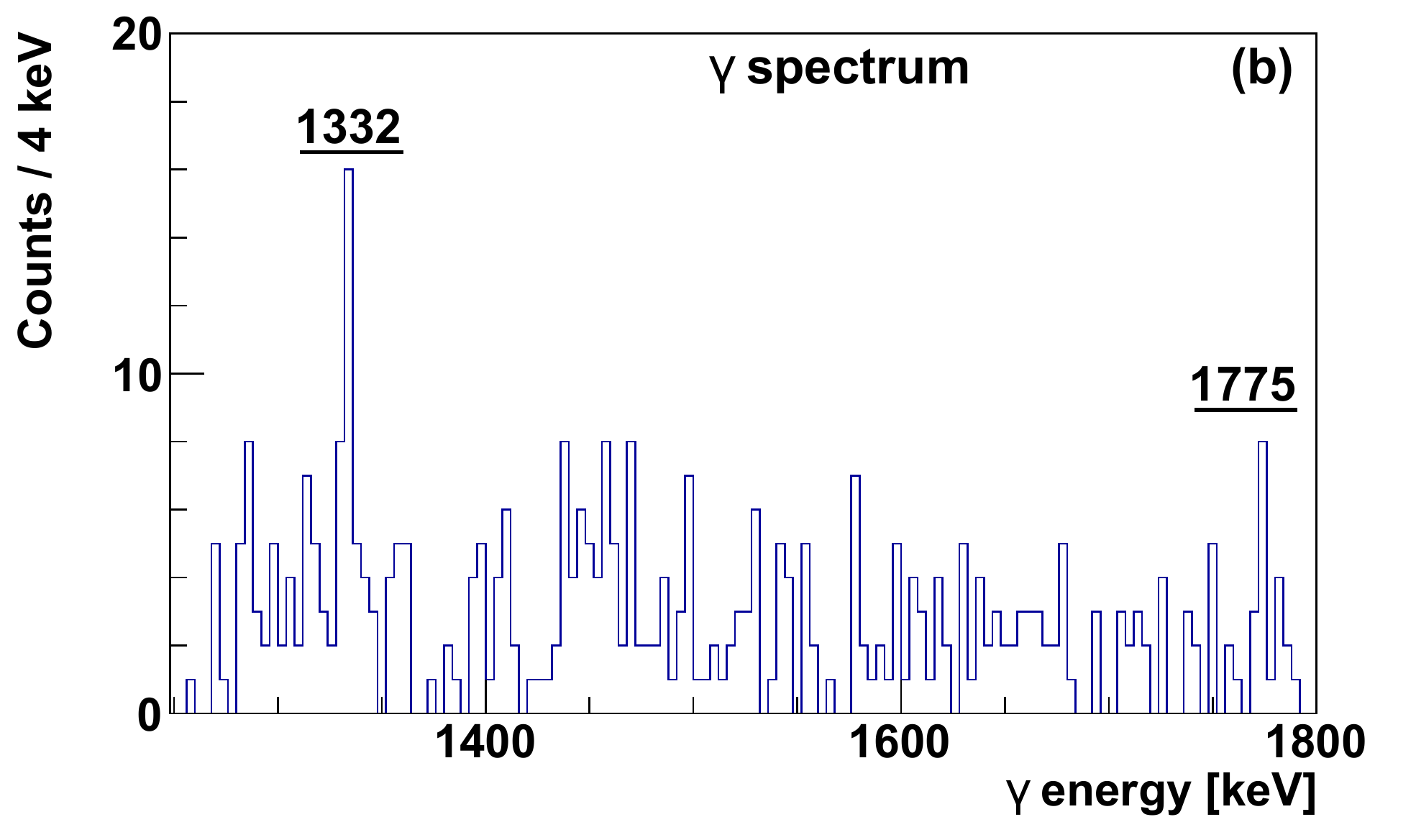}
	\end{minipage}
	\begin{minipage}{1.0\linewidth}
	  \includegraphics[width=1\columnwidth]{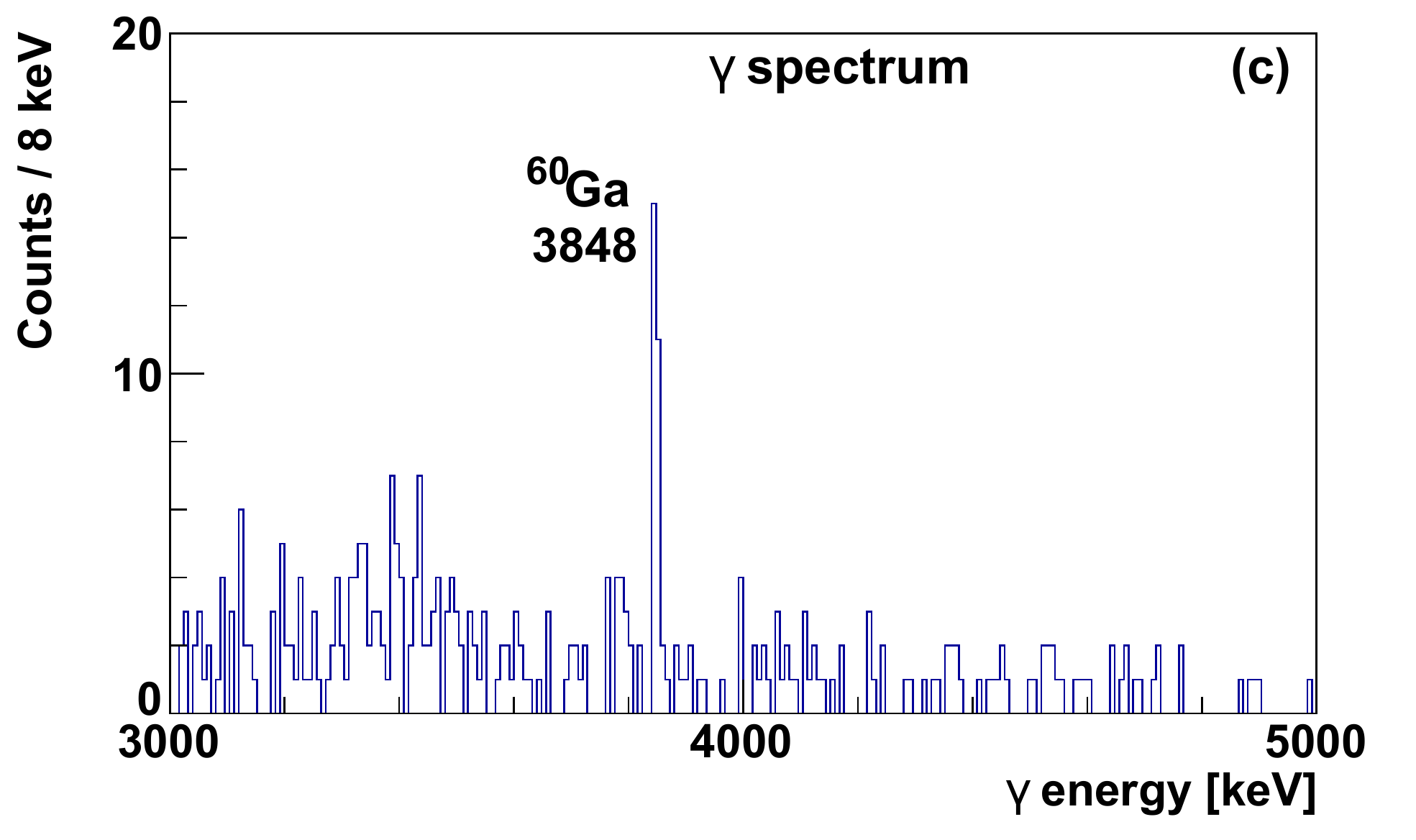}
	\end{minipage}
	\caption{$\gamma$-ray energy spectrum for decay events correlated with $^{60}$Ge implants. The observed $\gamma$ lines belong to different nuclei in the decay chain of $^{60}$Ge. Underlined values indicate $\gamma$ lines which are seen for the first time in the present work (see the text). Known $\gamma$ lines are labelled according to the parent nucleus. (a) The region [400,1100] keV is shown with bins of 4 keV/channel. (b) The region [1250,1800] keV is shown with bins of 4 keV/channel. (c) The region [3000,5000] keV is shown with bins of 8 keV/channel.}
	\label{gamma60}
\end{figure}

Fig. \ref{gamma60} shows the $\gamma$-ray energy spectrum from the $\beta$ decay of $^{60}$Ge ($T_z=\text{-}2$) to $^{60}$Ga ($T_z=\text{-}1$). Besides the 511 keV $\gamma$ line associated with the annihilation of the positrons emitted in the $\beta$ decay, seven $\gamma$ lines are observed at 463, 491, 837, 1004, 1332, 1775 and 3848 keV. These $\gamma$ transitions belong to different nuclei in the decay chain of $^{60}$Ge, as we discuss in detail below. The $\gamma$ lines which we observe for the first time are underlined in the figure. In particular, the $\gamma$ lines at 837, 1332 and 1775 keV are associated with the $\gamma$ de-excitation of states in the $^{60}$Ga daughter. The energies and intensities of all the $\gamma$ rays observed in Fig. \ref{gamma60} are given in Table \ref{protons}. The intensities $I_{\gamma}^{i}$ are obtained as:

\begin{equation}
  I_{\gamma}^{i} = \frac{N_{\gamma}^{i}}{\epsilon_{\gamma}^{i}~\epsilon_{decay}~N_{imp}~(1-\tau)} \,,
  \label{Eq2}
\end{equation}
where $N_{\gamma}^{i}$ represents the number of counts obtained from the integral of the $i^{th}$ $\gamma$ peak and $\epsilon_{\gamma}^{i}$ is the corresponding efficiency for $\gamma$ detection (Section \ref{exp}). The data acquisition system was triggered by either an implantation event or a decay event and the $\gamma$ rays were not included in the trigger. They were acquired in coincidence with decay events ($\beta$ or $\beta$p) and thus affected by the decay detection efficiency $\epsilon_{decay}$. This efficiency was estimated as in Ref. \cite{PhysRevC.93.044336} and it was 58(2)\% for $\gamma$ rays coincident with pure $\beta$ events, and 1 for the 463 keV $\gamma$ ray coincident with $\beta$p events.

\begin{table}[!t]
	\caption{Results for the $\beta^{+}$ decay of $^{60}$Ge. The first two columns show information on the $\beta$-delayed proton emission: the decay energies $E_p$ and their intensities $I_p$ (normalized to 100 decays). The last two columns report the $\gamma$-ray energies $E_{\gamma}$ and their intensities $I_{\gamma}$ (normalized to 100 decays of $^{60}$Ge). The observed $\gamma$ lines belong to different nuclei in the decay chain of $^{60}$Ge (see the text).}
	\label{protons}
	\centering
	\begin{ruledtabular} 
	  \begin{tabular}{l l | l l}
		\multicolumn{2}{l|}{$\beta$-delayed proton emission} & \multicolumn{2}{l}{$\beta$-delayed $\gamma$ de-excitation} \\		
 		  $E_p$(keV)   &    $I_p$(\%)    &  $E_{\gamma}$(keV)   &  $I_{\gamma}$(\%)  \\ \hline
			820(13)      &    2.8(4)       &   463.3(1)           &    8(1)            \\
			1076(23)     &    4.0(5)       &   491.2(2)           &    7(1)            \\
			1359(19)     &    5.1(4)       &   837.2(2)           &    9(1)            \\
			1684(17)     &    4.2(3)       &   1003.3(2)          &   15(2)            \\
			2067(15)$^a$  &    10.2(5)      &   1332.3(4)          &    4(1)            \\
			2522(15)$^b$  &    33(1)        &   1774.6(9)          &    2(1)            \\
			2981(23)     &    3.2(3)       &   3848.0(9)          &   13(3)            \\
			3490(22)      &    1.9(2)       &                      &                    \\
		\end{tabular}
	\end{ruledtabular}
\raggedright{$^a$ IAS proton emission to the first excited state of $^{59}$Zn. \\ $^b$ IAS proton emission to the g.s. of $^{59}$Zn.}
\end{table}

\begin{figure}[htb]
  \centering
	\includegraphics[width=1\columnwidth]{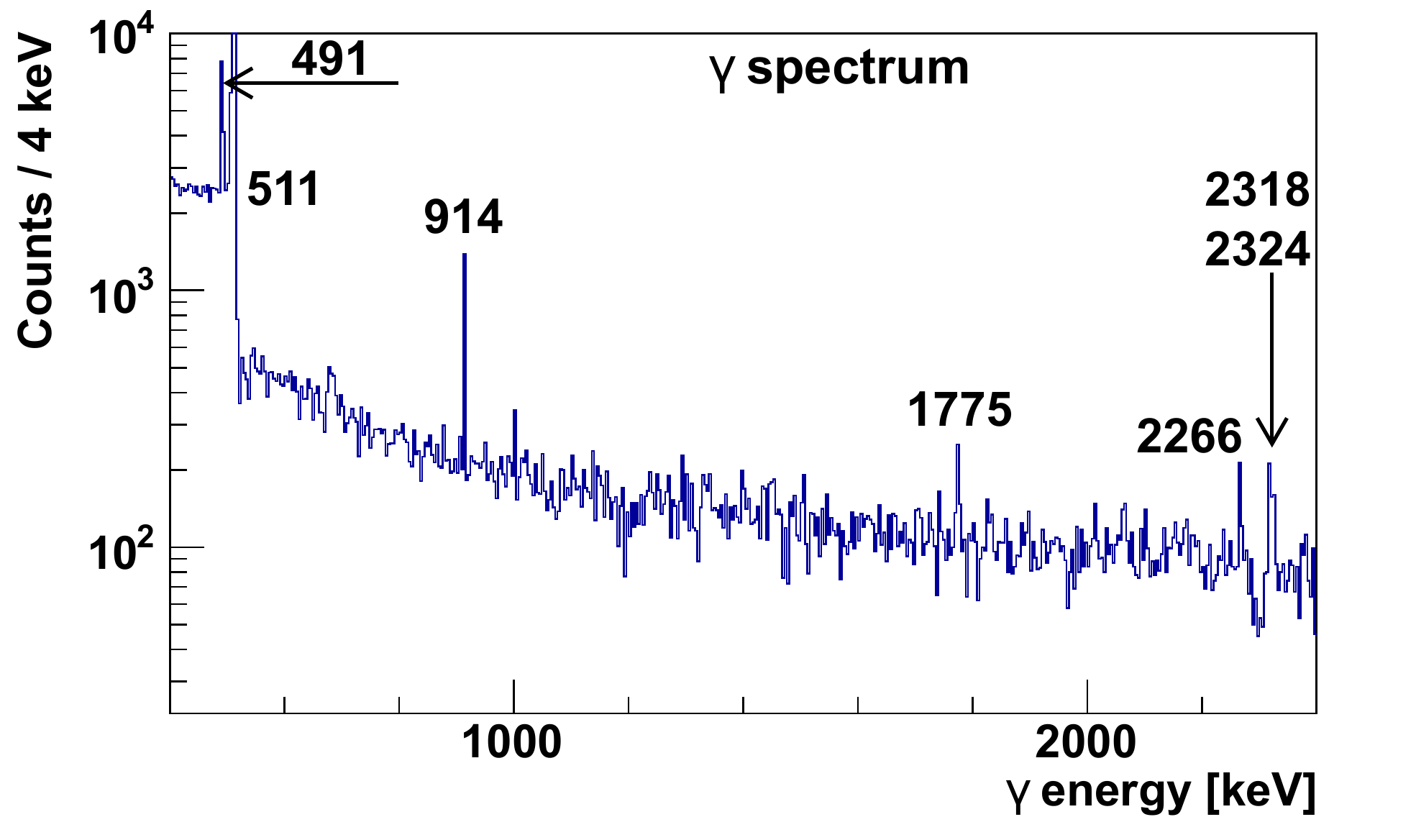}
	\caption{$\gamma$-ray energy spectrum for decay events correlated with $^{59}$Zn implants, showing the $\gamma$ de-excitation in $^{59}$Cu. $^{59}$Zn is also populated in the $\beta$-delayed proton decay of $^{60}$Ge.}
	\label{gamma59Zn}
\end{figure}

\begin{figure}[htb]
  \centering
	\includegraphics[width=1\columnwidth]{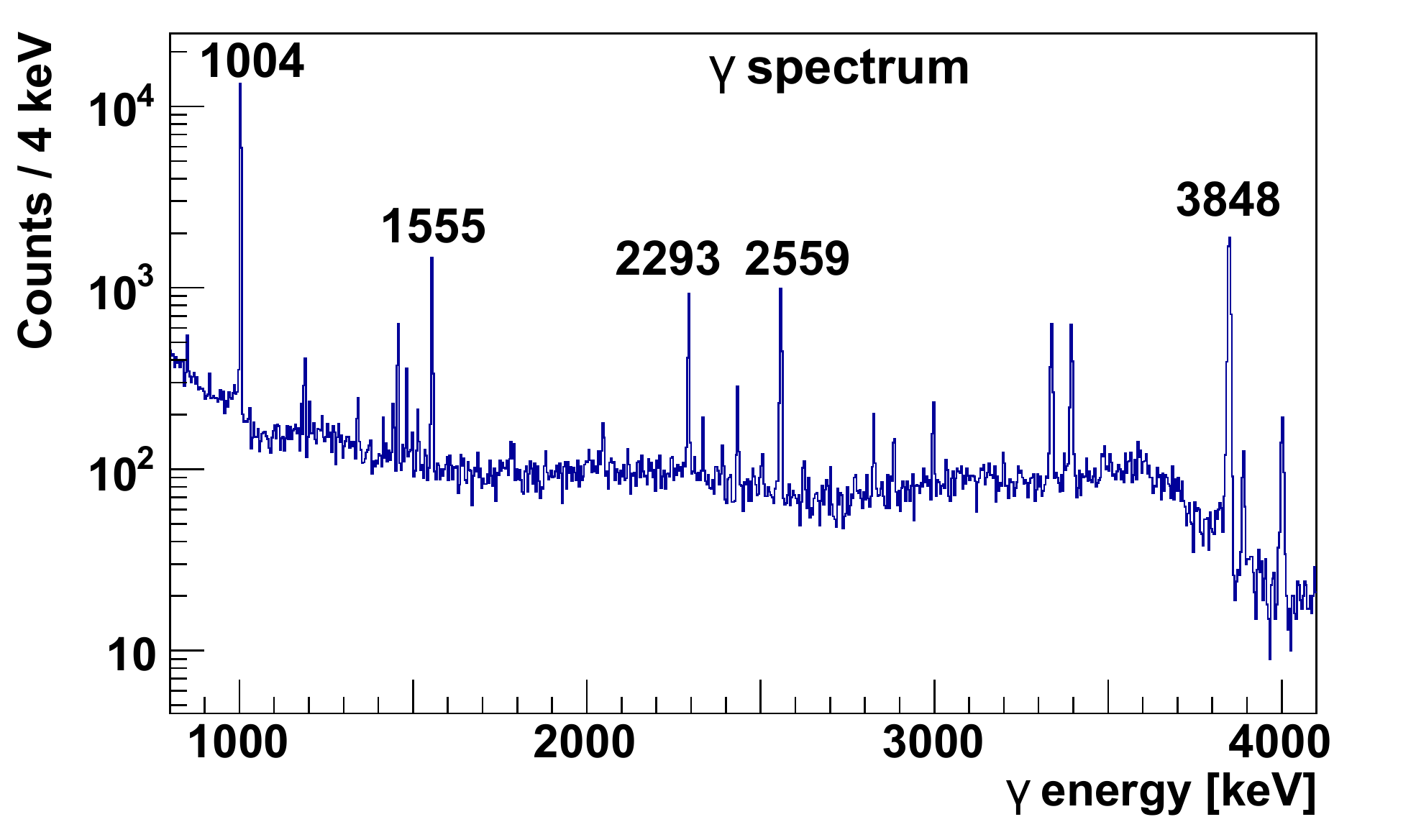}
	\caption{$\gamma$-ray energy spectrum for decay events correlated with $^{60}$Ga implants, showing the $\gamma$ de-excitation in $^{60}$Zn. The $^{60}$Ga nucleus is also populated in the $\beta$ decay of $^{60}$Ge.}
	\label{gamma60Ga}
\end{figure}

\begin{table}[htb]
	\caption{$\gamma$ rays observed in the $\beta^{+}$ decay of $^{60}$Ga. The $\gamma$-ray energies $E_{\gamma}$ and intensities $I_{\gamma}$ (normalized to 100 decays of $^{60}$Ga) from the present study are reported in the first two columns. For comparison, the last columns show $E_{\gamma}$ and $I_{\gamma}$ from Ref. \cite{Mazzocchi2001}, where the fourth column shows $I_{\gamma}$ relative to the 1004 keV $\gamma$ line, while the fifth column shows absolute $I_{\gamma}$ for 100 decays of $^{60}$Ga (calculated using our absolute intensity for the 1004 keV $\gamma$ ray).}
	\label{60GaGammas}
	\centering
	\begin{ruledtabular} 
	  \begin{tabular}{l l l l l}
 		  $E_{\gamma}$(keV)  & $I_{\gamma}$(\%) & $E_{\gamma}$(keV)$^a$  &  $I_{\gamma}^{rel}$  $^{ab}$   &  $I_{\gamma}$(\%)$^{ac}$ \\ \hline
			669.3(3)           &   0.36(9)        &                        &                         &           \\
			850.8(2)           &   0.75(10)       &                        &                         &           \\
			913.9(3)           &   0.3(1)         &                        &                         &           \\
			1003.5(1)         &   62(3)          &      1003.7(2)         &         100(17)         &   62(11)  \\
			1028.6(3)          &   0.38(8)        &                        &                         &           \\
			1188.4(2)          &   1.6(1)         &                        &                         &           \\
			1201.8(3)          &   0.29(7)        &                        &                         &           \\
			1413.7(3)          &   0.35(7)        &                        &                         &           \\
			1442.1(3)          &   0.40(8)        &                        &                         &           \\
			1481.4(3)          &   1.3(1)         &                        &                         &           \\
			1554.7(3)         &   7.0(5)         &      1554.9(6)         &          12(5)          &   7.4(31) \\
			1780.8(7)          &   0.2(1)         &                        &                         &           \\
			2047.2(6)          &   0.7(2)         &                        &                         &           \\
			2293.2(4)         &   6.3(5)         &      2293.0(10)        &          10(5)          &   6.2(31) \\
			2334.2(5)          &   0.8(2)         &                        &                         &           \\
			2434.2(4)          &   1.8(2)         &                        &                         &           \\
			2558.7(4)         &   8.5(6)         &      2559.0(8)         &          13(5)          &   8.1(31) \\
			2624.3(7)          &   0.3(1)         &                        &                         &           \\
			2826.0(5)          &   1.3(2)         &                        &                         &           \\
			2884.0(5)          &   0.8(2)         &                        &                         &           \\
			2996.8(5)          &   2.0(3)         &                        &                         &           \\
			3337.4(4)          &   7.1(6)         &                        &                         &           \\
			3394.8(4)          &   7.0(6)         &                        &                         &           \\
			3848.5(4)         &   38(3)          &      3848.3(7)         &          57(13)         &    35(8)  \\
			3889.1(5)          &   2.8(8)         &                        &                         &           \\
			4000.9(5)          &   2.8(4)         &                        &                         &           \\
			4805.0(5)          &   0.4(1)         &                        &                         &           \\
			4850.2(6)          &   0.2(1)         &                        &                         &           \\
			4891.9(5)          &   0.4(1)         &                        &                         &           \\                        
	  \end{tabular}
	\end{ruledtabular}
	\raggedright{$^a$ From Ref. \cite{Mazzocchi2001}. \\ $^b$ Relative intensity. \\ $^c$ Absolute intensity (calculated using our measured intensity for the 1004 keV $\gamma$ ray).}
\end{table}

\begin{figure}[!h]
  \centering
	\includegraphics[width=1\columnwidth]{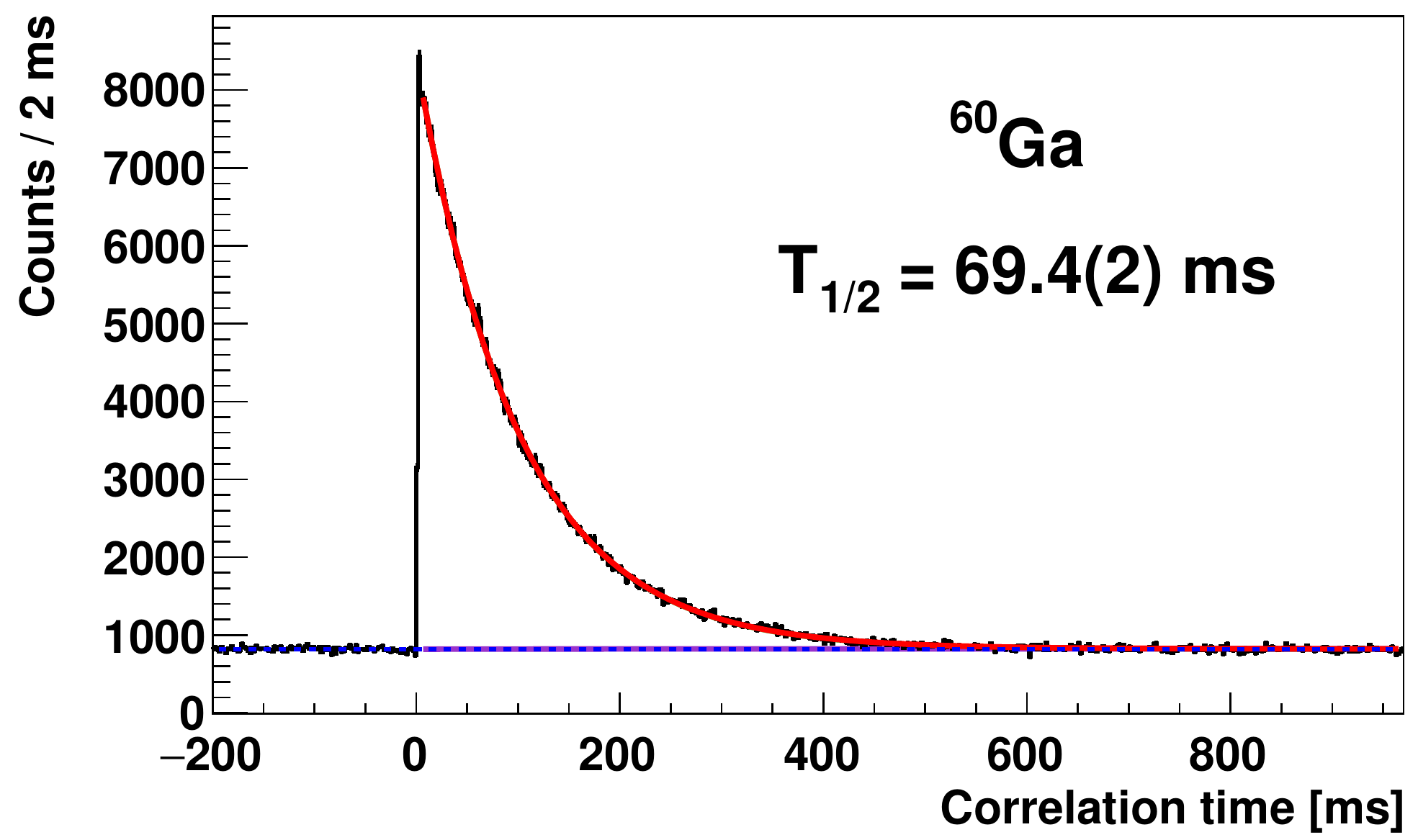}
 	\caption{Fit of the correlation-time spectrum obtained for the $\beta$ decay of $^{60}$Ga (red line). The fit includes the parent activity (green line practically coincident with the red line), the daughter activity (purple line) and the background (dashed blue line) fixed by fitting the backward-correlations part.}
  \label{T60a}
\end{figure}

\begin{figure*}[htb]
  \centering
	\includegraphics[height=1.0\textheight]{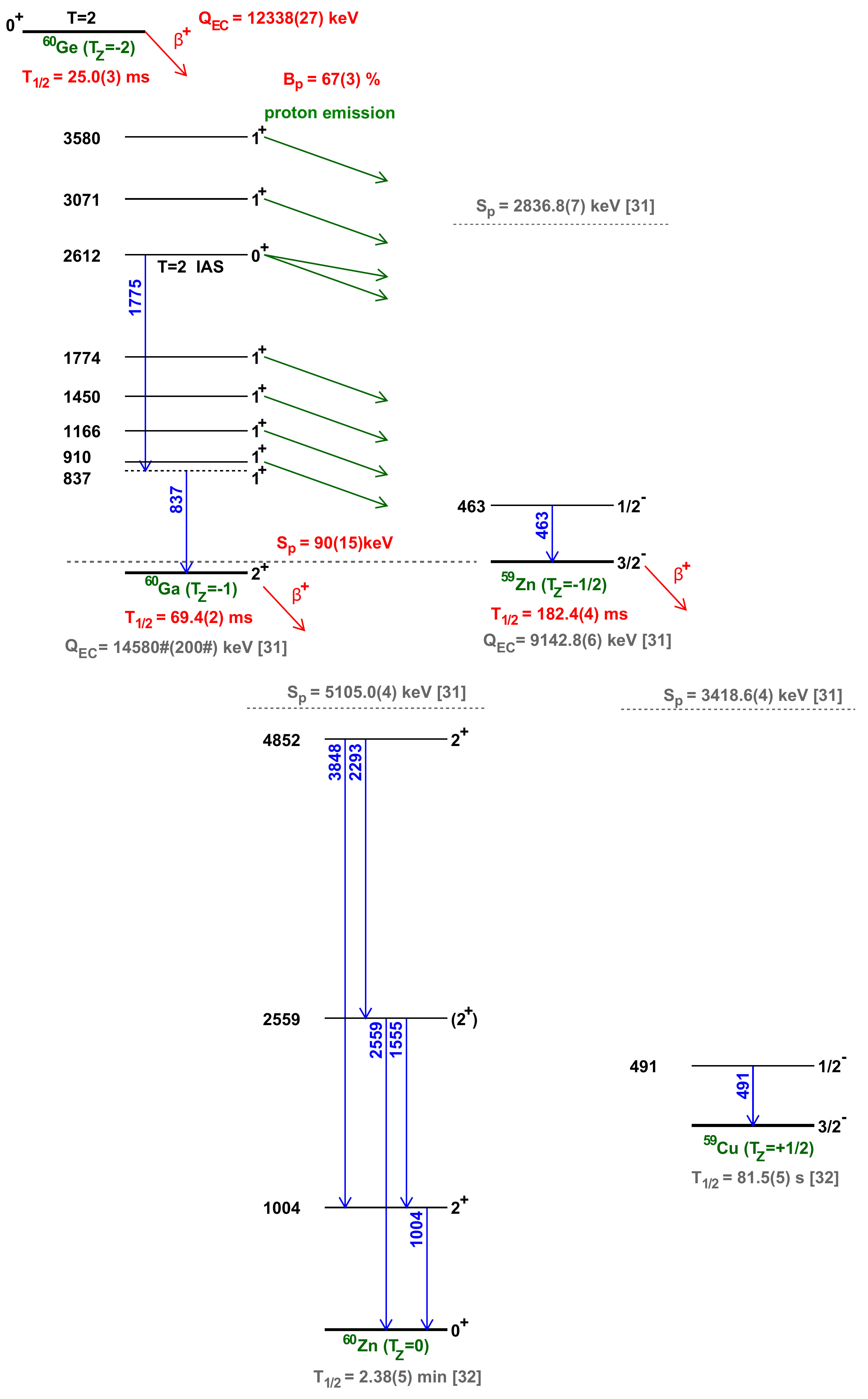}
 	\caption{Partial decay scheme of $^{60}$Ge deduced from the results of the present experiment. Evidence is seen of population of levels in four nuclei in the decay chain of $^{60}$Ge: $^{60}$Ga, $^{60}$Zn, $^{59}$Zn and $^{59}$Cu (see the text for more details). The observed $\gamma$ transitions are indicated by blue arrows, while green arrows indicate the observed proton emission. The quantities in red are deduced from the present data.}
  \label{decay60}
\end{figure*}

The reader may refer to Fig. \ref{decay60} for a summary of the decays that intervene in the following discussions. Below we discuss in detail the origin of all the observed $\gamma$ lines. As shown in Fig. \ref{dssd60}, $\beta$-delayed proton emission takes place in the $\beta$ decay of $^{60}$Ge. This process, passing through proton-unbound excited states of $^{60}$Ga, populates states in $^{59}$Zn \mbox{($T_z=\text{-}1/2$)}. $^{59}$Zn then $\beta$ decays to its mirror nucleus, $^{59}$Cu ($T_z=\text{+}1/2$) with \mbox{$T_{1/2}$ = 182.4(4)} ms (see below). The 491 keV $\gamma$ line is known and is the transition connecting the 1/2$^-$ first excited state of $^{59}$Cu to its 3/2$^-$ ground state (g.s.) \cite{HONKANEN1981,ARAI1984}.

Analysing the direct implantation of $^{59}$Zn (2.5$\times$10$^6$ implants) we obtained \mbox{$T_{1/2}$ = 182.4(4)} ms, that may be compared with the previous values of 210(20) \cite{HONKANEN1981}, 182.0(18) \cite{ARAI1984}, 173(14) \cite{Lopez2002} and 174(2) \cite{Kucuk2017} ms. We also obtained the $\gamma$-ray energy spectrum correlated with $^{59}$Zn implants (shown in Fig. \ref{gamma59Zn}) where, as expected, we observe known $\gamma$ lines de-exciting levels in $^{59}$Cu. Comparing the $\gamma$ lines of Figs. \ref{gamma60} and \ref{gamma59Zn}, the new $\gamma$ lines (463, 837 and 1332 keV) are not present in Fig. \ref{gamma59Zn}, while in both figures there is a line at 1775 keV. A known 1775-keV $\gamma$ ray de-excites the 2267 level in the $^{59}$Cu daughter nucleus, populating the 491 keV level. We think that the 1775 keV line of Fig. \ref{gamma60} belongs to a different transition, because otherwise we should have also observed the more intense 914 keV line in Fig. \ref{gamma60}. The proton separation energy in $^{59}$Zn is 2836.8(7) keV \cite{Wang2017} and the total $\beta$-delayed proton-emission branching ratio is 0.023(8)\% \cite{HONKANEN1981}, hence the contribution of the proton emission from $^{59}$Zn is negligible in the decay chain of $^{60}$Ge.

As mentioned above, the first excited state in $^{59}$Cu is 1/2$^-$ with 491 keV excitation energy and de-excites to its 3/2$^-$ g.s. through a $\gamma$ transition of the same energy \cite{HONKANEN1981,ARAI1984}. The g.s. of $^{59}$Zn is also 3/2$^-$ \cite{HONKANEN1981} but the energy of the expected 1/2$^-$ excited state is not known. The isopin symmetry and the expected similarity between $^{59}$Cu and $^{59}$Zn suggest that the 463 keV $\gamma$ line of Fig. \ref{gamma60} may be associated with the transition between the expected 1/2$^-$ excited state of $^{59}$Zn and its 3/2$^-$ g.s. An additional hint at this conclusion comes from the energy difference between the proton peaks at $E_p$ = 2067(15) and 2522(15) keV (Table \ref{protons}) which agrees, within the uncertainty, with the value of 463 keV. This suggests that the two proton peaks are indeed a doublet corresponding to the transitions from the IAS in $^{60}$Ga to the first excited and ground states of $^{59}$Zn, respectively. It should be noted that the parent of $^{59}$Zn, $^{59}$Ga, is probably unbound, therefore the $\beta$-delayed proton decay from $^{60}$Ge is presumably the best way to populate the first excited state at 463 keV in $^{59}$Zn. 

\begin{table}[!t]
	\caption{Mass excesses (in keV) of the four members of the $A$ = 60, $T$ = 2 mass multiplet that are used as the input for the IMME calculations. Mass excess (in keV) of the \mbox{$T_z$ = -2}, $^{60}$Ge$_{g.s.}$ nucleus obtained by the IMME calculations and compared with the 2016 and 2003 AME systematics \cite{Audi2017,Audi2003}.}
	\label{imme}
	\centering
	\begin{ruledtabular}
	  \begin{tabular}{c c c c}
		\multicolumn{4}{c}{$A$ = 60, $T$ = 2 input values} \\
		$T_z$ = +2  & $T_z$ = +1 & $T_z$ = 0  & $T_z$ = -1  \\
		-64473.1(4)$^a$ & -55804(5)$^a$  & -46807(24)$^a$ & -37405(15)$^b$ \\
		\cline{1-4}
		\multicolumn{2}{c}{IMME result for $T_z$ = -2} & 2016 AME \cite{Audi2017} & 2003 AME \cite{Audi2003} \\
		\multicolumn{2}{c}{-27678(22)} & -27090$^\#$(300$^\#$) & -27770$^\#$(230$^\#$) \\
		\end{tabular}
	\end{ruledtabular}
	\raggedright{$^a$ From Ref. \cite{Audi2017}. \\ $^b$ From the present $\beta$-delayed proton emission data. \\ $^\#$ Values obtained from systematics.}
\end{table}

On the other hand, the $\beta$ decay of $^{60}$Ge populates directly states in the $^{60}$Ga daughter. The latter then $\beta$ decays to $^{60}$Zn (with $T_z = 0$ and being its own mirror since $Z=N=30$) populating its IAS at 4852 keV, which de-excites mainly by the cascade of 3848 and 1004 keV $\gamma$ rays \cite{Mazzocchi2001} that are both observed in Fig. \ref{gamma60}. Looking at the direct implantation of $^{60}$Ga (7.6$\times$10$^5$ implants), we obtained the $\gamma$-ray energy spectrum of Fig. \ref{gamma60Ga}, showing the $\gamma$ de-excitation in $^{60}$Zn. In this spectrum we observe all the most intense $\gamma$ transitions at 1004, 1555, 2293, 2559 and 3848 keV, already seen in Ref. \cite{Mazzocchi2001}, and several other $\gamma$ lines listed in Table \ref{60GaGammas}. The intensities of the five known $\gamma$ transitions agree with those observed in Ref. \cite{Mazzocchi2001}. None of the new $\gamma$ lines (837, 1332 and 1775 keV) of Fig. \ref{gamma60} is observed in Fig. \ref{gamma60Ga}. A value \mbox{$B_p$ = 1.6(7)\%} was obtained for the $\beta$-delayed proton emission from $^{60}$Ga \cite{Mazzocchi2001}, populating $^{59}$Cu. Thus this process could also make a minor contribution to the observed 491 keV $\gamma$ line (Fig. \ref{gamma60}), even if we do not see this line in Fig. \ref{gamma60Ga}. Analysing the charged-particle spectrum for decay events correlated with $^{60}$Ga implants, we found the contribution of the $\beta$-delayed proton emission from $^{60}$Ga to be of the order of 1\% in comparison to that from $^{60}$Ge (Fig. \ref{dssd60}). Last but not least, we have improved our knowledge of the $^{60}$Ga half-life, obtaining $T_{1/2}$ = 69.4(2) ms from the fit shown in Fig. \ref{T60a}, that may be compared with the previous values of 70(15) \cite{Mazzocchi2001} and 76(3) ms \cite{Kucuk2017}.

The mass of $^{59}$Zn is known \cite{Audi2017}, but the mass of $^{60}$Ga has not been measured, which makes difficult to estimate properly the excitation energy $E_X$ of the IAS and other nuclear levels populated in the $^{60}$Ga daughter based on the observation of the proton peaks. The mass excess of $^{60}$Ge has also not been measured. From systematics ($^\#$) the most recent atomic mass evaluation (AME) \cite{Audi2017} gives a g.s. mass excess of \mbox{-27090$^\#$(300$^\#$)} keV for $^{60}$Ge and \mbox{-39590$^\#$(200$^\#$)} keV for $^{60}$Ga. From these masses the value $Q_{\beta}$ = 12500$^\#$(360$^\#$) keV is obtained \cite{Wang2017}. The proton separation energy in $^{60}$Ga is also not known experimentally. Systematics gives us $S_p$ = -340$^\#$(200$^\#$) keV \cite{Wang2017}, while a semiempirical estimate gives 40(70) keV \cite{Mazzocchi2001}. A systematic value for the excitation energy of the IAS is \mbox{$E_X^{IAS}$ = 2540$^\#$(50$^\#$)} keV \cite{Audi2017}. Ref. \cite{ciemny2016} calculated a value of 2520(280) keV based on Coulomb energy difference.

From our experimental data we can obtain the mass excess of the IAS in $^{60}$Ga. The information from the $\beta$-delayed proton emission gives us a proton energy \mbox{$E_p^{IAS}$ = 2522(15)} keV (Table \ref{protons}), remembering that \mbox{$E_X = E_p + S_p$}. Using our $E_p^{IAS}$ and the measured value for the mass excess of the $^{59}$Zn$_{g.s.}$, -47215.6(8) keV \cite{Audi2017}, one can determine the mass excess of $^{60}$Ga$^{IAS}$ being \mbox{-37405(15)} keV. Considering the latter mass together with the measured masses for the other three members of the $T$ = 2 isospin multiplet ($^{60}$Zn$^{IAS}$, $^{60}$Cu$^{IAS}$ and $^{60}$Ni$_{g.s.}$, listed in Table \ref{imme}), one can apply the Isobaric Multiplet Mass Equation (IMME) \cite{IMMEpaper,IMME} to determine the $^{60}$Ge$_{g.s.}$ mass excess, obtaining a value of -27678(22) keV. The 2003 AME value of \mbox{-27770$^\#$(230$^\#$)} keV \cite{Audi2003} agrees with our value better than the most recent AME mass excess. In Section \ref{ge62} we observe similar behaviour for the case of $^{62}$Ge. Issues with AMEs subsequent to that in 2003 have been reported already for proton-rich nuclei in this region of the mass chart \cite{PhysRevLett.112.222501,DelSanto2014453,PhysRevC.93.044336} and we believe that more mass measurements would be very useful to constrain the future AME in this mass region. 

The IAS in $^{60}$Cu, which is the mirror nucleus of $^{60}$Ga, lies at 2536.0(6) keV and de-excites through a $\gamma$ cascade with $\gamma$ energies 1866 and 670 keV \cite{NDS2013}. Based on mirror symmetry, together with the observed intensities for the 837 and 1775 keV $\gamma$ rays (Table \ref{protons}), we think that the IAS in $^{60}$Ga de-excites through a $\gamma$ cascade consisting of the 1775 and 837 keV $\gamma$ rays, in this order. From the energies of these $\gamma$ rays we can determine the excitation energy of the $^{60}$Ga$^{IAS}$, \mbox{$E_X^{IAS}$ = 2611.8(9)} keV, which agrees with the expected values \cite{Audi2017,ciemny2016}. We can also deduce other properties of $^{60}$Ga, such as the g.s. mass excess, -40016(15) keV, and $S_p$ = 90(15) keV. Finally, we calculate the value \mbox{$Q_{\beta}$ = 12338(27)} keV for the $\beta^+$ decay of $^{60}$Ge. 

\begin{table}[!t]
	\caption{Summary of the results for the $\beta^{+}$ decay of $^{60}$Ge. Level excitation energies $E_X$ in $^{60}$Ga, $\beta$ feedings $I_{\beta}$, Fermi $B$(F) and Gamow-Teller $B$(GT) transition strengths.}
	\label{table60}
	\centering
	\begin{ruledtabular} 
	  \begin{tabular}{l l l l}
 		  $E_X$(keV)         &  $I_\beta$(\%)  & $B$(F) &  $B$(GT)  \\ \hline
				3580(27)         &     1.9(2)      &        &  0.14(1)  \\
				3071(28)         &     3.2(3)      &        &  0.18(2)  \\
				2611.8(9)$^a$    &     45.3(20)    & 3.1(1) &           \\
				1774(23)         &     4.2(3)      &        &  0.11(1)  \\
				1450(25)         &     5.1(4)      &        &  0.11(1)  \\
				1166(28)         &     4.0(5)      &        &  0.074(9) \\
				910(20)          &     2.8(4)      &        &  0.044(6) \\
				837.2(2)         &     7(2)        &        &  0.11(3)  \\
		\end{tabular}
	\end{ruledtabular}
\raggedright{$^a$ IAS.}
\end{table}

The quantities deduced above are used to determine the excitation energies $E_X$ of the levels in $^{60}$Ga, the Fermi $B$(F) and Gamow-Teller $B$(GT) transition strengths. They are reported in Table \ref{table60} together with the $\beta$ feeding $I_{\beta}$ to each level, inferred from the intensities measured for both the $\beta$-delayed proton and $\gamma$ emissions (Table \ref{protons}). The level scheme deduced from the results of the present experiment is shown in Fig. \ref{decay60}.

We obtain a value $B$(F) = 3.1(1) that is smaller than the expected \mbox{$B$(F) = $|N-Z|$ = 4}. From Table \ref{table60}, \mbox{$I_\beta^{IAS}$ = 45.3\%} but we would need a feeding of 58\% in order to get the correct Fermi strength. A possible explanation could be that we are missing weak $\gamma$ branches de-exciting the IAS. As a matter of fact, we do not have enough information to place the 1332 keV $\gamma$ ray in the level scheme. If it de-excites the IAS, this latter would gain an additional feeding of 4\% (see Table \ref{protons}) out of the missing 12.7\%.

Another possible scenario would be to invert the order of the $\gamma$ rays in the cascade de-exciting the IAS, i.e., first the 837 and then the 1775 keV $\gamma$ rays. In this way the IAS would gain an additional feeding of 7\% and we would obtain a higher value for the strength, \mbox{$B$(F) = 3.6(2)}. According to this possibility, the 837 keV $\gamma$ ray would populate the level at $E_X$ = 1774 keV, which we already observed from the proton emission, but the balance of feeding populating and de-exciting this level would be negative (-3.0(15) \%). The 837 keV level would be removed from the decay scheme of Fig. \ref{decay60} and for this reason we indicated it with a dashed line. If this second scenario is confirmed, the emission of the 837 keV $\gamma$ ray followed by the emission of the proton from the 1774 keV level would be a new observation of $\beta$-delayed $\gamma$-proton decay, an exotic decay mode that we already observed in $^{56}$Zn \cite{PhysRevLett.112.222501}. A new experiment with higher statistics would be desirable to shed light on these aspects.

In the decay of the $T_z=\text{-}2$ nuclei the $\beta$-delayed proton emission from the IAS populated in the daughter nucleus is isospin-forbidden, but usually it is observed and this is attributed to a $T$ = 1 isospin impurity in the IAS wave function \cite{PhysRevLett.112.222501,PhysRevC.93.044336}. As in other \mbox{$T_z=\text{-}2$} cases, in the decay of $^{60}$Ge we observe competition between the $\beta$-delayed proton and $\gamma$ emissions from the IAS populated in $^{60}$Ga. However here the proton emission is found to be dominant, being 95\% of the observed IAS de-excitation (74.5\% if we account for the missing $\gamma$ feeding). The explanation for this behaviour may lie in nuclear structure reasons.

\section{\label{ge62}Beta decay of $^{62}$G\MakeLowercase{e}}

$^{62}$Ge is a $T_z=\text{-}1$ proton-rich nucleus which was only poorly known at the time of the present experiment. The high-intensity $^{78}$Kr beam available at RIKEN made it possible to record the unprecedented statistics of 2.1$\times$10$^6$ implants of $^{62}$Ge.

\begin{figure}[!t]
  \centering
	\includegraphics[width=1\columnwidth]{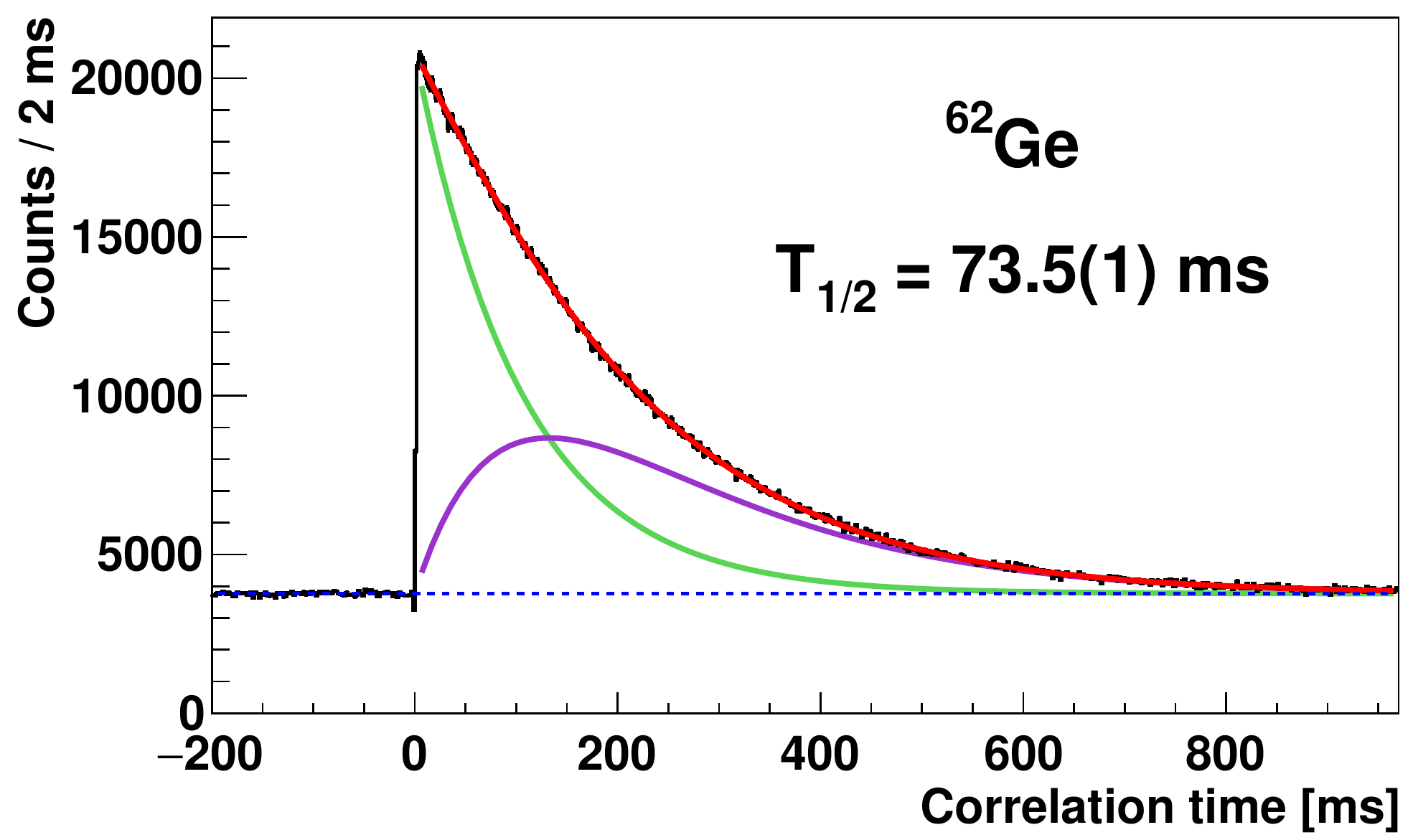}
 	\caption{Fit of the correlation-time spectrum obtained for the $\beta$ decay of $^{62}$Ge. The green line indicates the parent activity, the purple line is the daughter activity, the dashed blue line is the background (fixed by fitting the backward-correlations part of the spectrum) and the red line is the sum of all.}
  \label{T62}
\end{figure}

Fig. \ref{T62} shows the correlation-time spectrum obtained for the $\beta$ decay of $^{62}$Ge. The $^{62}$Ge half-life was determined by fitting the correlation-time spectrum using the Bateman equations \cite{Bateman1910} including the $\beta$ decay of $^{62}$Ge, the growth and decay of its daughter $^{62}$Ga (with \mbox{$T_{1/2}$ = 116.121(21)} ms \cite{Audi2017}) and the random correlation background. A half-life of $T_{1/2}$ = 73.5(1) ms is obtained. The maximum likelihood and least squares minimization methods gave the same result. This half-life value agrees with and improves the precision of our previous measurement (76(6) ms obtained with 6.1$\times$10$^3$ implants of $^{62}$Ge \cite{Kucuk2017}). As already discussed in Ref. \cite{Kucuk2017}, there is an older measurement, 129(35) ms \cite{Lopez2002}, lying inside two standard deviations mainly because of the poorer statistics. Additionally, another value of 82.9(1.4) ms is reported in Ref. \cite{Grodner2014} again with much poorer statistics.

\begin{figure}[!t]
  \centering
	\includegraphics[width=1\columnwidth]{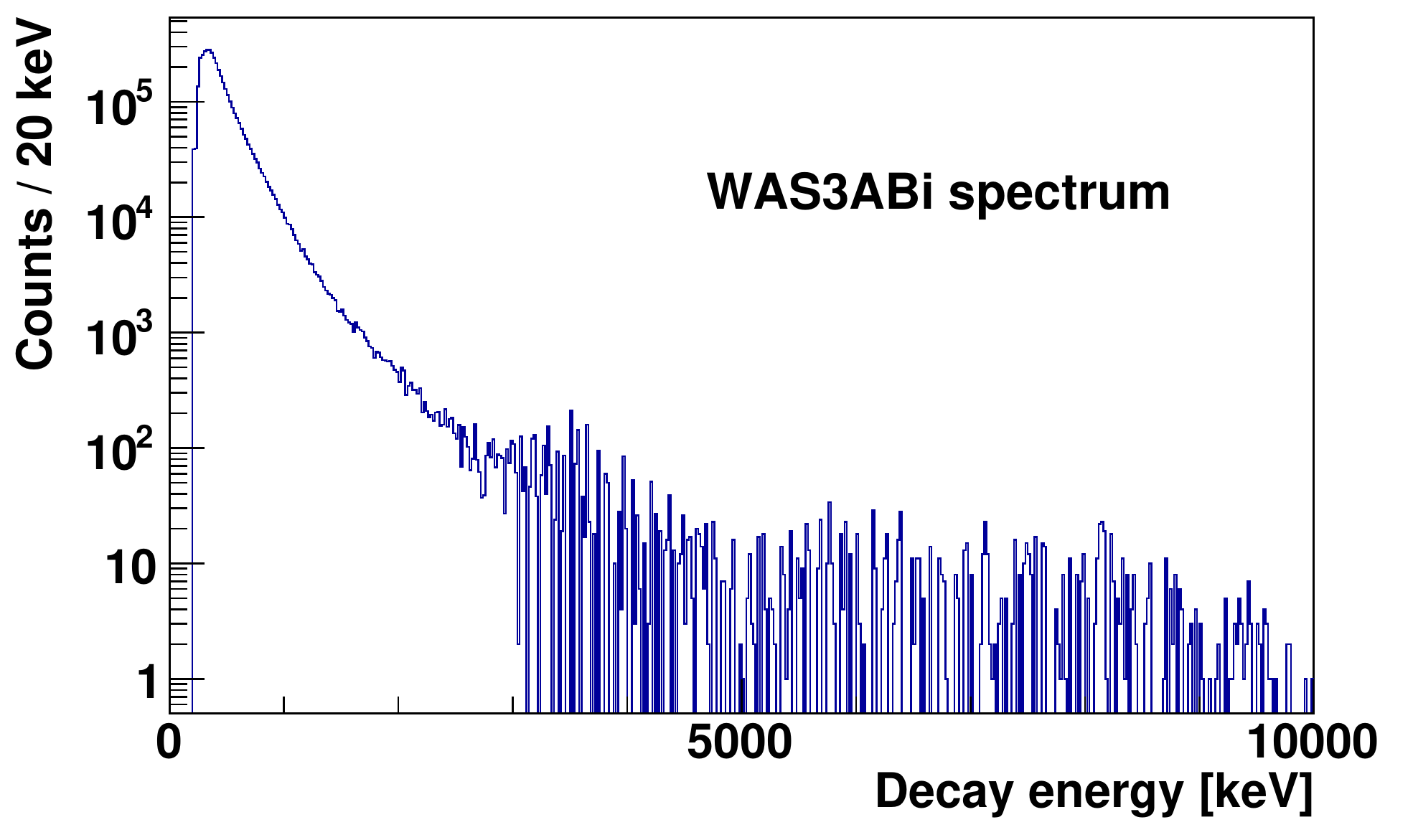}
 	\caption{Charged-particle energy spectrum for decay events correlated with $^{62}$Ge implants.}
  \label{dssd62}
\end{figure}

The charged-particle energy spectrum from the decay of $^{62}$Ge is shown in Fig. \ref{dssd62}. The bump at low energy is due to the detection of $\beta$ particles, while no discrete peaks from $\beta$-delayed proton emission are observed. The proton separation energy in the $^{62}$Ga daughter is 2927(16) keV \cite{Wang2017}, above this value proton emission is allowed. From our data we estimated an upper limit of 5\textperthousand $~$for the total proton-emission branching ratio.

\begin{figure}[!t]
	\begin{minipage}{1.0\linewidth}
    \centering
    \includegraphics[width=1\columnwidth]{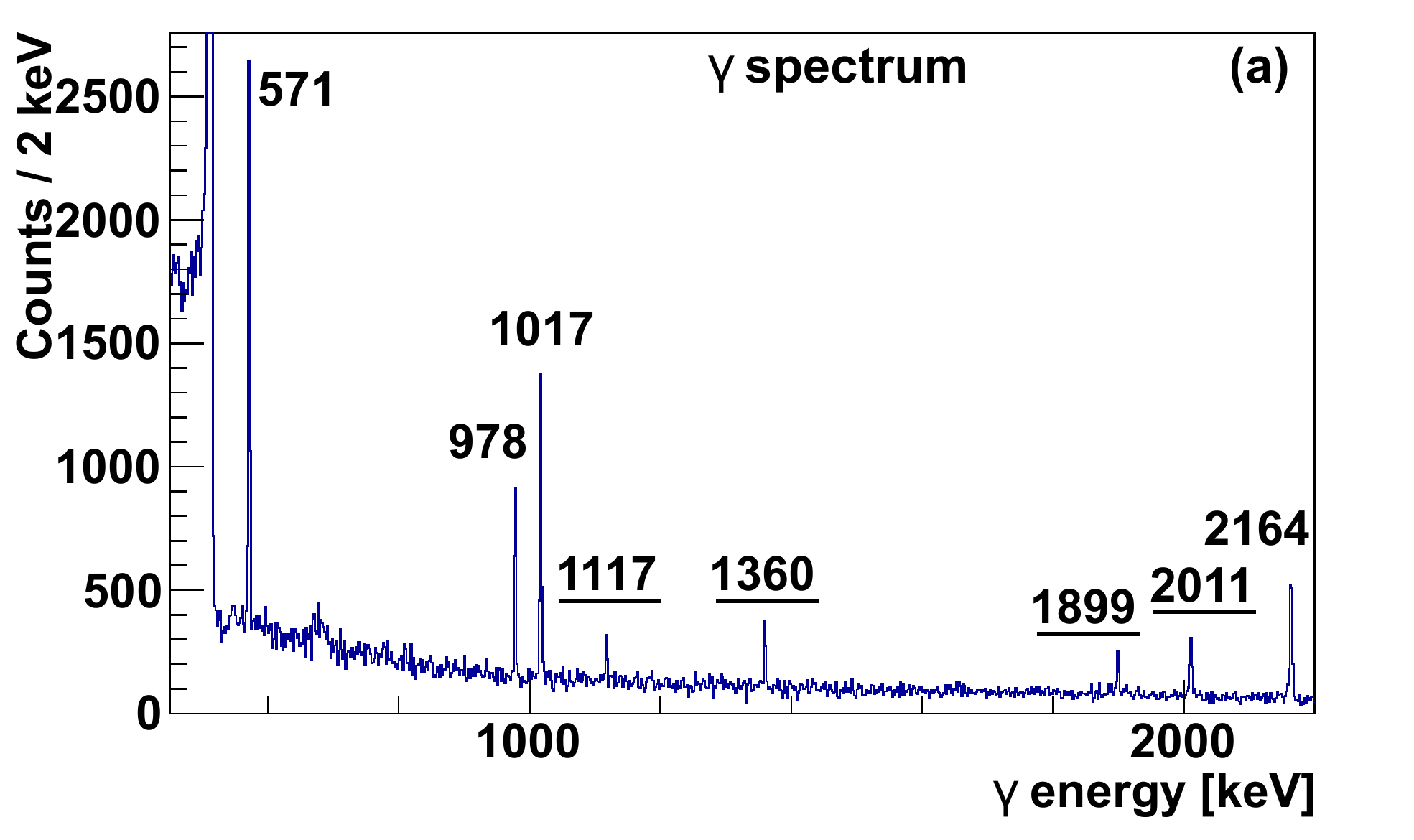}
	\end{minipage}
	\begin{minipage}{1.0\linewidth}
	  \includegraphics[width=1\columnwidth]{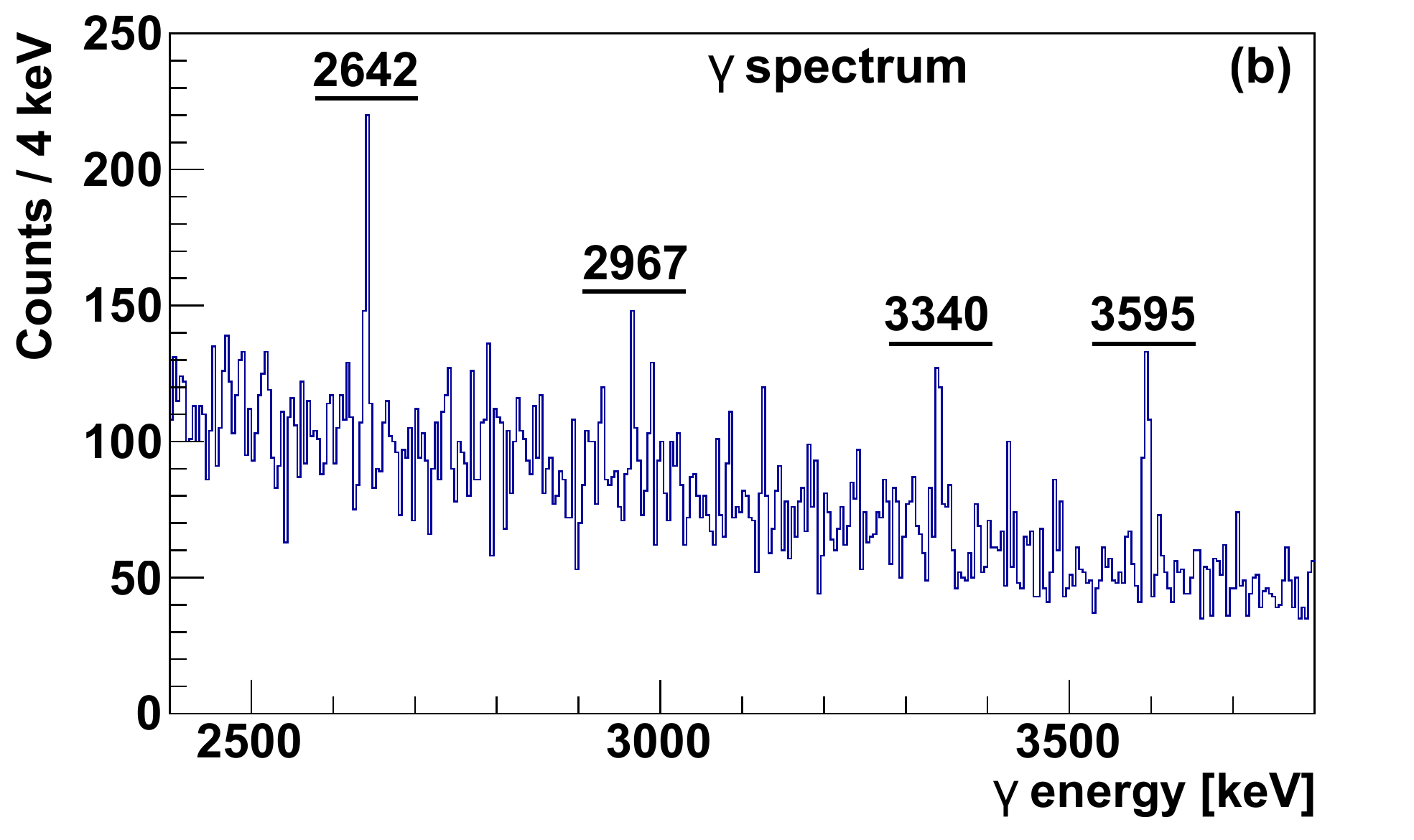}
	\end{minipage}
	\caption{$\gamma$-ray energy spectrum for decay events correlated with $^{62}$Ge implants. Underlined values indicate $\gamma$ lines which are seen for the first time in the present work. (a) The region [450,2200] keV is shown with bins of 2 keV/channel. (b) The region [2400,3800] keV is shown with bins of 4 keV/channel.}
	\label{gamma62}
\end{figure}

The $\gamma$-ray energy spectrum from the $\beta$ decay of $^{62}$Ge is shown in Fig. \ref{gamma62}. We observed twelve $\gamma$ lines at 571, 978,	1017,	1117,	1360,	1899,	2011,	2164,	2642,	2967,	3340	and 3595 keV. Eight of them were observed for the first time, while the four most intense lines (571, 978, 1017 and 2164 keV) were also seen in Ref. \cite{Grodner2014} together with two other $\gamma$ lines at 1247 and 2414 keV which we did not see. The $\gamma$ lines which we observed for the first time are underlined in Fig. \ref{gamma62}.

\begin{figure}[htb]
  \centering
	\includegraphics[width=1\columnwidth]{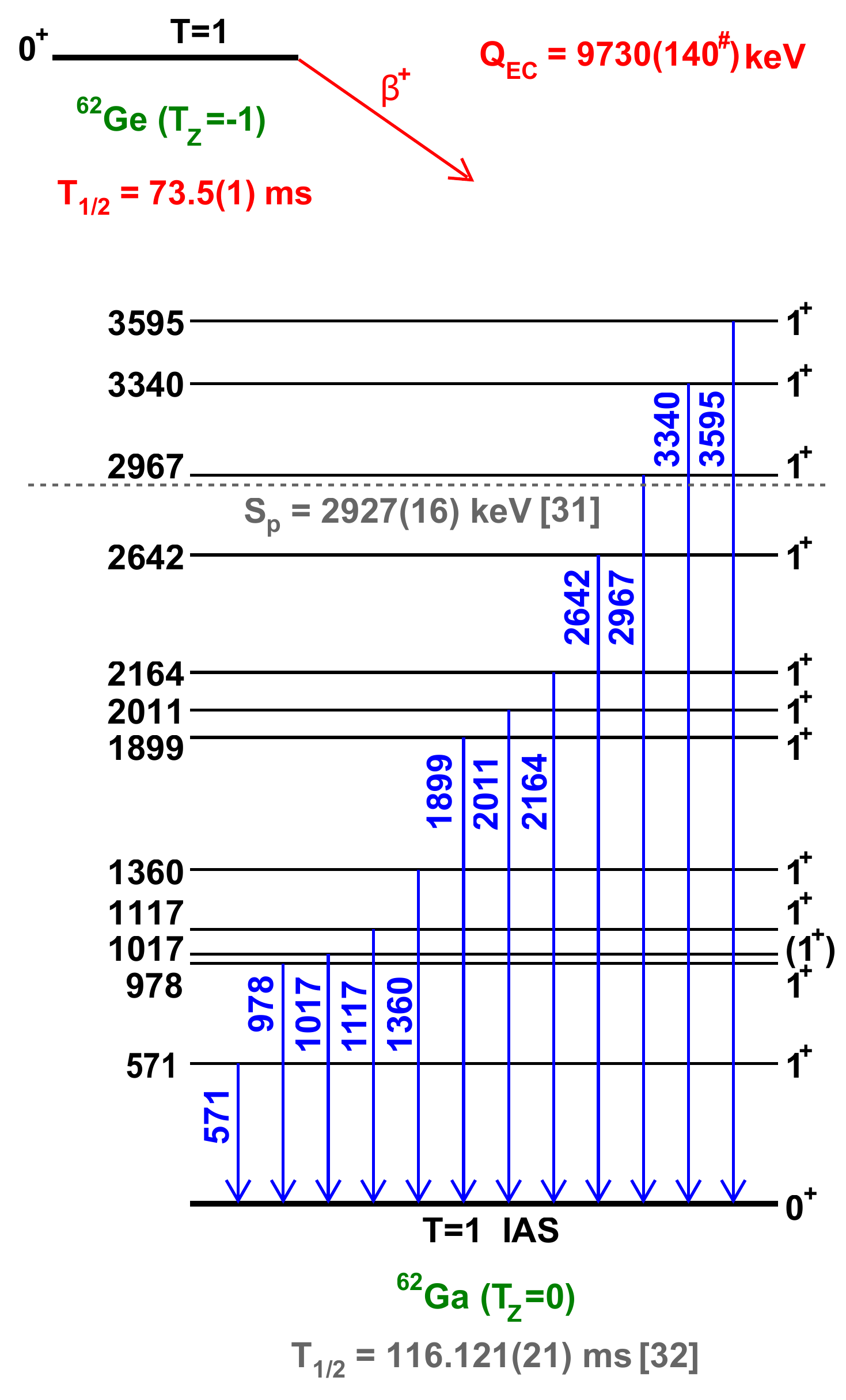}
 	\caption{Partial decay scheme of $^{62}$Ge deduced from the results of the present experiment. The observed $\gamma$ transitions are indicated by blue arrows. The quantities in red are deduced from the present data.}
  \label{decay62}
\end{figure}

The decay scheme of $^{62}$Ge deduced from the results of the present experiment is shown in Fig. \ref{decay62}. The $\beta^+$ decay of the \mbox{$J^{\pi}$ = 0$^+$}, $T$ = 1 g.s. of $^{62}$Ge \mbox{($T_z=\text{-}1$)} populates the 0$^+$, $T$ = 1 IAS in the $^{62}$Ga daughter ($T_z = 0$) through a super-allowed Fermi transition and 1$^+$, $T$ = 0 states through Gamow-Teller transitions. According to the $quasi$-$rule$ of Warburton and Weneser, namely that the M1 transitions between \mbox{$J^{\pi}$ = 1$^+$}, $T$ = 0 states are strongly suppressed \cite{Mompurgo58,Wilkinson69}, one can expect that all the populated 1$^+$, $T$ = 0 states decay predominantly to the 0$^+$, \mbox{$T$ = 1} g.s. of $^{62}$Ga. We have found evidence for this $quasi$-$rule$ in other $T_z=\text{-}1$ cases \cite{PhysRevC.91.014301}. In order to test this rule for the present case, we looked for small branches from the upper 1$^+$ states to the lower 1$^+$ levels. Even considering the most favourable case, i.e., a transition from the 1$^+$ state at 2164 keV to the first 1$^+$ state at 571 keV, the de-exciting $\gamma$ ray was not observed. We set an upper limit of 1 \textperthousand $~$on its intensity.

In a fusion-evaporation-reaction study of $^{62}$Ga \cite{Rudolph2004} a 2$^+$ level was proposed at 1017 keV based on the observation of a weak $\gamma$ ray at 446 keV in coincidence with the 571 keV $\gamma$ ray, with no observation of the 1017 keV $\gamma$ ray. Considering two following in-beam studies, Ref. \cite{David2013} also reported the 446 keV $\gamma$ ray, but Ref. \cite{Henry2015} did not observe it; moreover in both references the 1017 keV $\gamma$ ray was not seen. In the present $\beta$-decay experiment and also in those of Ref. \cite{Grodner2014} an intense $\gamma$ ray was seen at 1017 keV, while the 446 keV $\gamma$ ray was not observed. Hence we think that the level we see at 1017 keV is not the same level as in Ref. \cite{Rudolph2004} and therefore it is a 1$^+$ state, like the other levels populated in the decay. This is further supported by the non-observation of weak $\gamma$ branches from the upper excited 1$^+$ levels to the 1017 keV level which, if it were a 2$^+$ level, would not be affected by the Warburton and Weneser $quasi$-$rule$.

Since there is no appreciable $\beta$-delayed proton emission from $^{62}$Ge, Eq. \ref{Eq2} is simplified and the intensity $I_\gamma^{i}$ of each $\beta$-delayed $\gamma$ ray can be calculated as follows:

\begin{equation}
  I_{\gamma}^{i} = \frac{N_{\gamma}^{i}}{\epsilon_{\gamma}^{i}~N_{\beta}} \,,
  \label{Eq3}
\end{equation}
with the advantage of not depending on $\epsilon_{decay}$. Here $N_\beta$ represents the total number of $\beta$ events correlated with $^{62}$Ge and is obtained from the half-life fit in Fig. \ref{T62}.

\begin{table}[!t]
	\caption{Summary of the results for the $\beta^{+}$ decay of $^{62}$Ge. The first two columns show the $\gamma$-ray energies $E_{\gamma}$ and their intensities $I_{\gamma}$ (normalized to 100 decays). The last four columns give the level excitation energies $E_X$ in $^{62}$Ga, $\beta$ feedings $I_{\beta}$, Fermi $B$(F) and Gamow-Teller $B$(GT) transition strengths to the $^{62}$Ga levels. The $B$(GT) values are calculated by imposing \mbox{$B$(F) = 2.0} (see the discussion in the text).}
	\label{table62}
	\centering
	\begin{ruledtabular}
	  \begin{tabular}{l l l l l l}
		 $E_{\gamma}$(keV) & $I_{\gamma}$(\%) & $E_X$(keV) & $I_\beta$(\%) & $B$(F) & $B$(GT)   \\ \hline
     3594.7(5)	  &   0.6(1)   	& 3594.7(5)	   & 0.6(1)	  & & 0.07(1)	            \\ 
     3339.6(5)	  &   0.30(6)	  & 3339.6(5)	   & 0.30(6)	& & 0.030(7)	          \\	
     2966.8(5)	  &   0.35(6)	  & 2966.8(5)	   & 0.35(6)	& & 0.028(5)	          \\		
     2641.8(5)	  &   0.4(1)	  & 2641.8(5)	   & 0.4(1)	  & & 0.029(7)	          \\		
     2164.1(4)	  &   2.6(2)	  & 2164.1(4) 	 & 2.6(2)	  & & 0.13(1)	 	          \\	
     2010.9(4)	  &   0.96(8)	  & 2010.9(4)	   & 0.96(8)	& & 0.045(5)	          \\		
     1899.3(4)	  &   0.58(6)	  & 1899.3(4)	   & 0.58(6)	& & 0.025(3)	          \\		
     1359.7(2)	  &   0.70(5)	  & 1359.7(2)    & 0.70(5)	& & 0.022(2)	          \\		
     1117.4(2)	  &   0.41(4)	  & 1117.4(2)	   & 0.41(4)	& & 0.011(2)	          \\		
     1017.1(1)  	&   2.6(1)	  & 1017.1(1)	   & 2.6(1)	  & & 0.067(6)	          \\		
     978.3(1)	    &   1.8(1)	  & 978.3(1)	   & 1.8(1)	  & & 0.047(4)	          \\		
     571.3(1)	    &   3.4(1)	  & 571.3(1)	   & 3.4(1)	  & & 0.068(6)	          \\		
		              &             &  g.s.$^a$    & 85.3(3)$^b$  & 2.0 &             \\	
	  \end{tabular}
	\end{ruledtabular}
	\raggedright{$^a$ IAS. \\ $^b$ The g.s. to g.s. feeding is $I_\beta^{IAS}=(100-\Sigma_{i}I_{\gamma}^i)$.}
\end{table}

Table \ref{table62} summarises our results for $^{62}$Ge. The $\gamma$-ray energies $E_{\gamma}$ and intensities $I_{\gamma}$ are reported in the first two columns of the table. Assuming that all the populated 1$^+$ states decay directly to the 0$^+$ g.s. of $^{62}$Ga (as discussed above), the excitation energies $E_X$ of the levels in $^{62}$Ga are deduced directly from the corresponding $E_{\gamma}$. Similarly, the $\beta$ feeding $I_{\beta}$ to each level is inferred from the respective $I_{\gamma}$. The third and fourth columns of Table \ref{table62} give $E_X$ and $I_{\beta}$, respectively. Finally, the last two columns give the $B$(F) and $B$(GT) transition strengths, where the $B$(GT) values are obtained by setting \mbox{$B$(F) = 2.0} (see the discussion below).

The Fermi transition connects the $^{62}$Ge$_{g.s.}$ to its IAS, the $^{62}$Ga$_{g.s.}$. The IAS $\beta$ feeding (g.s.-to-g.s.) is calculated by subtracting all the indirect $\gamma$ feeding to the $^{62}$Ga$_{g.s.}$, \mbox{$I_\beta^{IAS}=(100-\Sigma_{i}I_{\gamma}^i)$ = 85.3(3)\%}. Then, in order to calculate the $\beta$-decay strengths, one needs to know the mass excesses of the nuclear states involved in the transition. The mass excess of the $^{62}$Ga$_{g.s.}$ is known experimentally and has a value of -51986.9(6) keV \cite{Audi2017}. However the g.s. mass of $^{62}$Ge is not known. Since the \mbox{$A$ = 62}, \mbox{$T$ = 1} isospin multiplet is a triplet ($^{62}$Ge, $^{62}$Ga and $^{62}$Zn), we cannot use the IMME \cite{IMMEpaper,IMME} to determine the $^{62}$Ge$_{g.s.}$ mass, hence we have to rely on systematics. The most recent AME gives a value of -41740$^\#$(140$^\#$) keV \cite{Audi2017}, which gives a Fermi strength of 1.5(1), i.e., much smaller than the expected \mbox{$B$(F) = $|N-Z|$ = 2}. If we make the Fermi strength equal to 2, then we obtain a mass excess of -42258 keV for the $^{62}$Ge$_{g.s.}$ and, assuming for this mass a conservative uncertainty of 140 keV as in the mass evaluation systematics, we obtain the value $Q_{\beta}$ = 9730(140$^\#$) keV shown in Fig. \ref{decay62}. The 2003 AME value of -42240$^\#$(140$^\#$) keV \cite{Audi2003} is in agreement with our deduced mass. In addition to the cases of $^{62}$Ge and $^{60}$Ge (Table \ref{imme}), similar issues with the AMEs subsequent to the 2003 AME have been already reported for proton-rich nuclei in this region of the mass chart \cite{PhysRevLett.112.222501,DelSanto2014453,PhysRevC.93.044336}. More mass measurements are required in this region to constrain the future AME.

From our deduced g.s. mass excesses for $^{60}$Ge and $^{62}$Ge we can extract the two-neutron separation energy in $^{62}$Ge, $S_{2n}$ = 30723(142) keV. It should be noted that the values from both the 2016 (30970$^\#$(330$^\#$) keV \cite{Wang2017}) and 2003 (30620$^\#$(270$^\#$) keV \cite{AUDI2003337}) AME systematics agree well with our value.

The $B$(GT) values shown in Table \ref{table62} agree with those observed in Ref. \cite{Grodner2014}, confirming the absence of enhanced low-lying Gamow-Teller strength in $^{62}$Ga and, hence, the negligible role of the $T$ = 0 proton-neutron pairing condensate in $A$ = 62 \cite{Grodner2014}.

\section{\label{concl}Conclusions}

We have studied the $\beta$ decay of the very neutron deficient $^{60}$Ge and $^{62}$Ge isotopes. We measured half-lives of $T_{1/2}$ = 25.0(3) ms for $^{60}$Ge, 69.4(2) ms for the daughter $^{60}$Ga and 73.5(1) ms for $^{62}$Ge. In all cases we have improved the precision in comparison with values in the literature \cite{ciemny2016,Mazzocchi2001,Kucuk2017,Grodner2014}.

The decay of $^{60}$Ge proceeds mainly by $\beta$-delayed proton emission populating $^{59}$Zn with a total branching ratio of 67(3)\%. Through this process we have observed for the first time the 1/2$^-$ first excited state in $^{59}$Zn and determined its excitation energy as 463.3(1) keV. The $\beta$-delayed $\gamma$ emission is also seen in competition, populating energy levels previously unknown in the drip-line $^{60}$Ga nucleus. Three new $\gamma$ lines (837, 1332 and 1775 keV) have been observed and the level scheme deduced.

The decay of $^{62}$Ge proceeds mostly by direct population of the $^{62}$Ga ground state with a g.s. to g.s. feeding of 85.3(3)\%. A total of twelve $\gamma$ lines have been observed, eight of them for the first time. The Warburton and Weneser $quasi$-$rule$ \cite{Mompurgo58,Wilkinson69}, already observed in previous $T_z=\text{-}1$ cases \cite{PhysRevC.91.014301}, has again been confirmed here. Moreover, even if there are a few differences in the observed $\gamma$ rays with respect to Ref. \cite{Grodner2014}, our study confirms that there is no evidence of enhanced low-lying Gamow-Teller strength in $^{62}$Ga due to isoscalar proton-neutron pairing.

Finally, the g.s. mass excesses of $^{60}$Ge, $^{60}$Ga and $^{62}$Ge have been deduced and the absolute $\beta$-decay transition strengths have been determined.

\begin{acknowledgments}
This work was supported by the Spanish MICINN grants FPA2014-52823-C2-1-P and FPA2017-83946-C2-1-P (MCIU/AEI/FEDER); Centro de Excelencia Severo Ochoa del IFIC SEV-2014-0398; $Junta~para~la~Ampliaci\acute{o}n~de~Estudios$ Programme (CSIC JAE-Doc contract) co-financed by FSE; MCIU grant IJCI-2014-19172. We acknowledge the support of the Generalitat Valenciana Grant No. PROMETEO/2019/007. We acknowledge the support of STFC(UK) grant P005314/1 and ANID/CONICYT FONDECYT Regular 1171467. R.B.C. acknowledges support from the Max-Planck Partner group.
\end{acknowledgments}

\bibliography{references}

\end{document}